\documentclass[12pt]{spieman}  
\usepackage{amsmath,amsfonts,amssymb}
\usepackage{graphicx}
\usepackage{setspace}
\usepackage{tocloft}
\usepackage{lineno}
\usepackage{aas_macros}
\usepackage{hyperref}

\newcommand{\arcsec}{\mbox{$^{\prime \prime}$}}

\title{Polycyclic Aromatic Hydrocarbons (PAHs) in the High-redshift Universe: Prospect of the PRIMA FIRESS low-resolution spectroscopy}

\author[1,2,*]{Ilsang Yoon}
\author[3]{Brandon Hensley}
\author[4]{Thomas S. -Y. Lai}
\author[5]{Irene Shivaei}
\author[6]{Ismael Garc{\'\i}a-Bernete}
\author[7]{Grant P. Donnelly}
\author[8]{Alexandra Pope}
\author[7]{J. D. T. Smith}
\author[2]{Paul Torrey}

\affil[1]{National Radio Astronomy Observatory, 520 Edgemont Road, Charlottesville, VA 22903, USA}
\affil[2]{Department of Astronomy, University of Virginia, P.O. Box 3818, Charlottesville, VA 22903, USA}
\affil[3]{Jet Propulsion Laboratory, California Institute of Technology, 4800 Oak Grove Drive, Pasadena, CA 91109, USA}
\affil[4]{IPAC, California Institute of Technology, 1200 E. California Blvd., Pasadena, CA 91125}
\affil[5]{Centro de Astrobiolog\'ia (CAB), CSIC-INTA, Carretera de Ajalvir km 4, Torrej\'on de Ardoz, 28850, Madrid, Spain}
\affil[6]{Centro de Astrobiolog\'ia (CAB), CSIC-INTA, Camino Bajo del Castillo s/n, E-28692 Villanueva de la Can\~ada, Madrid, Spain}
\affil[7]{Ritter Astrophysical Research Center, University of Toledo, Toledo, OH 43606, USA}
\affil[8]{Department of Astronomy, University of Massachusetts, Amherst, MA 01003, USA}

\cftpagenumbersoff{figure}
\cftpagenumbersoff{table} 
\begin{document} 
\maketitle

\begin{abstract}
The integrated luminosity from the features of the polycyclic aromatic
hydrocarbons (PAHs) exceeds the luminosity from atomic and molecular emission lines in the star-forming regions in galaxies and is a potential tracer of galaxy-scale star formation and molecular gas content of the high-redshift universe. We simulate the observable PAH spectra using the PRobe far-Infrared Mission for Astrophysics far-infrared enhanced survey spectrometer (FIRESS) and investigate the capability of the FIRESS low-resolution spectroscopy for observing PAH emission spectrum from high-redshift galaxies. Our investigation suggests that (1) PRIMA observations of PAH emission are $\gtrsim10$ times more efficient at detecting galaxies than the VLA observations of CO(1-0) for galaxies with the same infrared luminosity, (2) PRIMA/FIRESS can detect the PAH emission from galaxies with $L_{\mbox{\tiny IR}}\sim10^{12}L_{\odot}$ up to the end of reionization (and possibly beyond, if $L_{\mbox{\tiny IR}}\sim10^{13}L_{\odot}$), (3) the PAH band ratios measured from a full spectral fitting and from a simple flux
"clipping" method are different and vary depending on the interstellar
radiation field strength, and (4) PRIMA/FIRESS can also be used as the PAH mapping instrument to measure star formation and redshift of the galaxies in high-redshift protoclusters.
\end{abstract}


{\noindent \footnotesize\textbf{*}Ilsang Yoon,  \linkable{iyoon@nrao.edu} }

\begin{spacing}{1}   

\section{Introduction}
\label{sec:intro}  
One of the key quests in studying galaxy formation and evolution is tracing cosmic star formation and the molecular gas content (i.e., fuel for star formation) as a function of redshift from the present to the epoch of the first galaxy formation. Different parts of the electromagnetic spectrum (emission lines and continuum) trace different `faces' of star formation (SF): the ultraviolet (UV) and optical spectrum trace the young stellar populations and surrounding ionized medium, while the radio and far-infrared (FIR) spectrum trace the dust, atoms, and molecules mainly heated by nearby SF (see Madau \& Dickinson 2014\cite{madau_dickinson_2014} for review of the cosmic star formation history).

Traditionally, carbon monoxide (CO) has been used as a tracer of molecular gas content because the CO molecule---the second most abundant (after H$_2$) molecule in a galaxy's interstellar medium (ISM)---is excited by collision with H$_2$ molecules (i.e., star-forming material) in star-forming molecular clouds. The Kennicutt-Schmidt (KS) relation\cite{kennicutt_1998} captures the connection between molecular gas and SF. As a result, the CO emission is known to be well correlated with the distribution of molecular gas and the IR luminosity (a tracer of SF) for galaxies in the nearby\cite{leroy_etal_2011,bigiel_etal_2011,schruba_etal_2012} and distant\cite{carilli_walter_2013,daddi_etal_2010,daddi_etal_2015,tacconi_etal_2010,tacconi_etal_2018,tacconi_etal_2020} Universe. Although, under the assumption of thermal equilibrium in the star-forming molecular gas, the increasing brightness of CO with higher $J$ rotational transitions is an advantage for observing high-redshift galaxies as demonstrated by the use of radio-millimeter-FIR observing facilities (e.g., VLA, ALMA, NOEMA), the lowest transition line, CO(1-0) with the $X_{\mbox{\tiny CO}}$ factor\cite{bolatto_etal_2013}(i.e., a conversion factor between CO line brightness and H$_{2}$ column density), is the best tracer of the total molecular gas mass.    

Another useful tracer of SF is the mid-IR (MIR) emission from Polycyclic Aromatic Hydrocarbon (PAH) molecules\cite{peeters_etal_2004,rigopoulou_etal_1999} which absorb UV and optical photons from star-forming regions and supply the bulk of the photoelectrons responsible for heating interstellar gas\cite{weingartner_draine_2001}. Although the AGN feedback in AGN-host galaxy also affects the PAH emission from a central $\approx100$ pc region \cite{garcia-bernete_etal_2022,garcia-bernete_etal_2024,jensen_etal_2017,lai_etal_2023,zhang_etal_2024}, the integrated PAH emission from $1-10$ kpc$^2$ scale in strongly star-forming galaxies can be as large as up to 20\% of the galaxy's total infrared luminosity\cite{smith_etal_2007}, which makes it a potentially efficient way of detecting galaxies in high-redshift as already suggested by the previous observations of the Spitzer space telescope \cite{li_2020}.

\begin{figure}
\begin{center}
\begin{tabular}{ccc}
\includegraphics[width=0.31\textwidth]{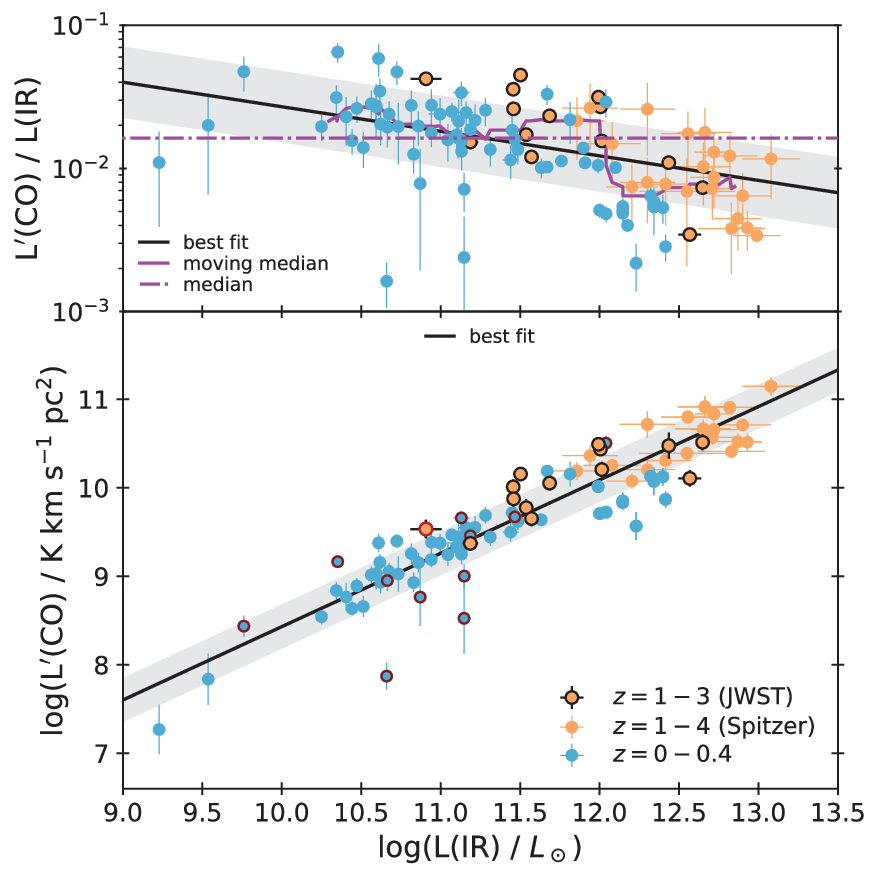} &
\includegraphics[width=0.31\textwidth]{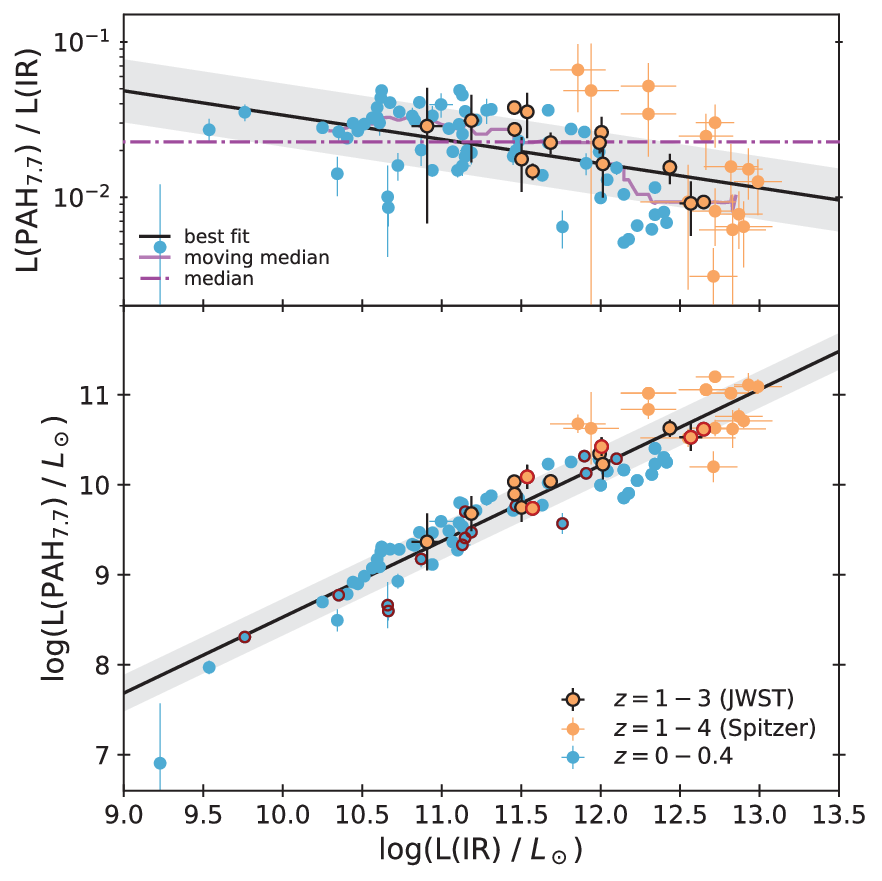} &
\includegraphics[width=0.35\textwidth]{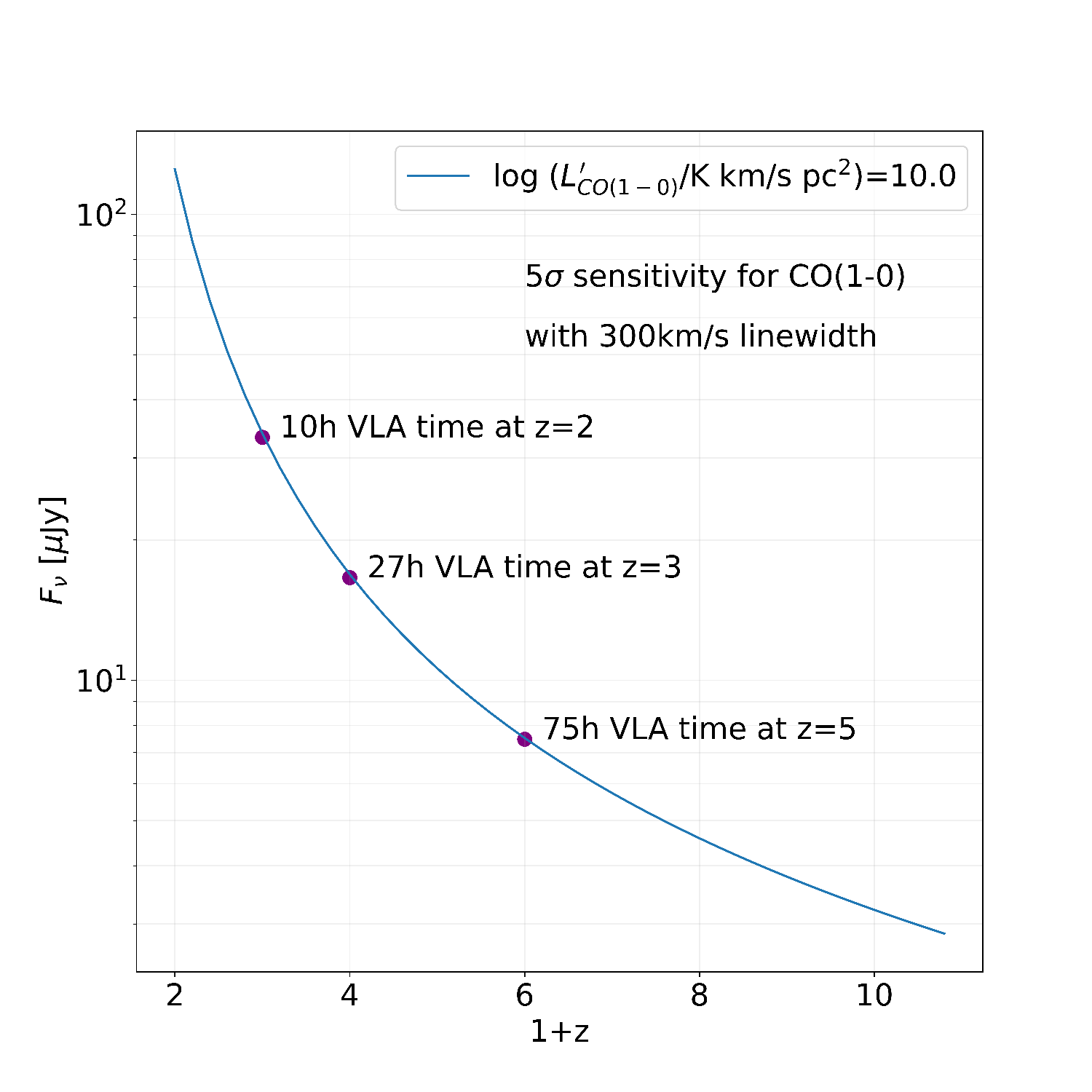} \\
\scriptsize (a) $L_{\mbox{\tiny IR}}$ correlation with CO & \scriptsize (b) $L_{\mbox{\tiny IR}}$ correlation with PAH & \scriptsize (c) VLA observing time   
\end{tabular}
\end{center}
\caption 
{\textit{Panel (a) and (b): }Correlation of $L_{\mbox{\tiny IR}}$ with CO(1-0) and with PAH$_{7.7}$ luminosity for nearby and distant galaxies (up to $z=4$). Figures are created using the data from Shivaei \& Boogaard 2024\cite{shivaei_boogaard_2024}. \textit{Panel (c): }VLA $5\sigma$ sensitivity with 300 km s$^{-1}$ bandwidth for a galaxy with $L^{\prime}_{\mbox{\tiny CO(1-0)}}=10^{10}$K km s$^{-1}$ pc$^{-2}$ to be detected in CO(1-0) as a function of redshift. The VLA observing time (including overhead) estimates are shown at $z$=2, 3, and 5.}\label{fig:co_pah_lir}
\end{figure} 

From the KS relation, one can expect a correlation between PAHs, CO molecules, and star formation rate (SFR). As shown in the previous literature, CO(1-0) luminosity obtained either from direct measurement or conversion from higher-$J$ rotational transitions, is correlated with the total IR luminosity (for example, Fig.~\ref{fig:co_pah_lir}(a)) that traces SFR. However, PAH emission is also well correlated with the total IR luminosity\cite{lin_etal_2024,shivaei_etal_2024} as shown in Fig.~\ref{fig:co_pah_lir}(b) as an example. The PAH emission correlates strongly with CO emission at sub-kiloparsec scales over diverse environments of local star-forming and active galactic nucleus (AGN) host galaxies\cite{gao_etal_2022,leroy_etal_2023,zhang_ho_2023}, with an indication of stronger correlation than that of CO–SFR or PAH–SFR\cite{whitcomb_etal_2023}. The PAH-CO connection is also seen beyond the local Universe\cite{pope_etal_2013,cortzen_etal_2019,shivaei_boogaard_2024}. Based on the compilations of large samples of galaxies with both mid-infrared spectroscopy and CO observations\cite{kirkpatrick_etal_2014,pope_etal_2013,smith_etal_2007}, a universal PAH–CO correlation is found from $z\sim0$ to $z\sim4$ on galaxy integrated scales\cite{cortzen_etal_2019}.

However, observing CO(1-0) becomes difficult with increasing redshift. To illustrate this difficulty, we estimate the VLA 5$\sigma$ sensitivity for CO(1-0) observation from a galaxy with $L^{\prime}_{\mbox{\tiny CO(1-0)}}=10^{10}$K km s$^{-1}$ pc$^{-2}$ (and the corresponding IR luminosity $L_{\mbox{\tiny IR}}=10^{12}$$L_{\odot}$) by assuming a linewidth for sensitivity estimate to be 300 km s$^{-1}$\cite{carilli_walter_2013}. Fig.~\ref{fig:co_pah_lir}(c) shows the $5\sigma$ RMS sensitivity for CO(1-0) as a function of the galaxy redshift. We also present the total VLA observing time (including calibration) for the corresponding sensitivity at $z=2, 3, 5$. It is clear that even at $z=3$, the observing time ($\approx30$ hours) is already significantly large and the CO(1-0) observation is not practically feasible if galaxies are at $z>4$. Furthermore, at high redshift, the CMB temperature ($T(z)=2.73\times(1+z)$) is comparable to or can be even higher than the typical CO excitation temperature ($T_{\mbox{\tiny CO(1-0)}}\sim10-20$K\cite{goldsmith_2013}). Thus, the contrast between the observed CO brightness and the CMB background decreases with redshift\cite{dacunha_etal_2013}, making the observing situation worse. Given the challenges in observing CO from high-redshift galaxies, one may consider the PAH emission as an alternative to CO to trace the star-forming galaxies across cosmic time: from cosmic noon ($z=2$--$4$) to cosmic dawn ($z\sim10$).

In the post-Spitzer era, JWST/MIRI Medium Resolution Spectroscopy (MRS) is the only instrument capable of observing PAH emission spectrum from high-redshift galaxies. It already has a showcase: 3.3$\mu$m PAH emission spectrum from a lensed star-forming galaxy at $z=4.22$\cite{spilker_etal_2023}, and the ongoing JWST/MIRI MRS programs will observe PAH emission spectra from star-forming galaxies (e.g., GO-5279, PI: Shivaei) and AGNs (e.g., GO-3158, PI: Mainieri) at cosmic noon. However, the dramatically decreasing sensitivity of JWST/MIRI in the long wavelength channels (channel 4 in MIRI/MRS) and the degraded performance of the MRS long wavelength channels due to a time-dependent evolution in the effective count rate\footnote{\linkable{https://www.stsci.edu/contents/news/jwst/2023/temporal-behavior-of-the-miri-reduced-count-rate}}, do not allow for probing PAH emission from galaxies at $z>4$. 

The next generation far-infrared space telescope, PRobe far-Infrared Mission for Astrophysics\footnote{\linkable{https://prima.ipac.caltech.edu/}} (PRIMA) is a 1.8 meter diameter, cryogenically-cooled to 4.5K, far-infrared observatory concept for the community in the 2030 decade, and includes both a sensitive wideband spectrometer, Far-InfraRed Enhanced Survey Spectrometer (FIRESS) and a multi-band spectrophotometric imager/polarimeter, PRIMA imager (PRIMAger).
The PRIMA/FIRESS is an ideal instrument to observe PAH emission spectrum from galaxies up to the end of reionization and possibly beyond, as proposed in, for example, chapters 18, 21, and 22 in the PRIMA GO science book\cite{prima_science_2023}. Moreover, the wavelength range and the sensitivity of PRIMA/FIRESS complement the JWST/MIRI MRS, enabling characterization of the spectrum with the full set of PAH bands for galaxies in the nearby and distant Universe. In this paper, we simulate an ensemble of the noise-added PAH spectra for PRIMA/FIRESS to observe galaxies in a wide range of redshift and different PAH properties, investigate the redshift limit for detection, and compare the PAH luminosity measurement from the full spectrum using all four bands and a segment of the spectrum from a single band.

\begin{table}[ht]
\caption{FIRESS Base Grating Module Instrument Parameters \cite{bradford_etal_2024}} 
\label{tab:instpar}
\begin{center}       
\begin{tabular}{|l|l|l|l|l|} 
\hline
\rule[-1ex]{0pt}{3.5ex}  Parameter & Band 1 & Band 2 & Band 3 & Band 4  \\
\hline\hline
\rule[-1ex]{0pt}{3.5ex}  Spectral Range ($\mu$m) & $24-43$ & $42-76$ & $74-134$ & $130-235$  \\
\hline
\rule[-1ex]{0pt}{3.5ex}  Spectral Sampling ($\mu$m) & $0.23$ & $0.41$ & $0.73$ & $1.29$  \\
\hline
\rule[-1ex]{0pt}{3.5ex}  Resolving Power & $90-150$ & $85-120$ & $90-125$ & $95-130$  \\
\hline
\rule[-1ex]{0pt}{3.5ex}  Pixel Size on Sky (arcsec) & $7.6$ & $7.6$ & $12.7$ & $22.9$ \\
\hline
\rule[-1ex]{0pt}{3.5ex}  Line Sensitivity{$^\dag$} (W/m$^{-2}$) & $1.9\times10^{-19}$ & $1.9\times10^{-19}$ & $1.9\times10^{-19}$ & $1.9\times10^{-19}$ \\
\hline
\rule[-1ex]{0pt}{3.5ex}  Continuum Sensitivity{$^\ddag$} ($\mu$Jy) & $64$ & $125$ & $196$ & $193$ \\
\hline
\end{tabular}
\end{center}
\footnotesize{$^\dag$ $5\sigma$ point source sensitivity in 1-hour integration for an unresolved line\\
$^\ddag$ $5\sigma$ point source sensitivity in 1-hour integration for $\mbox{R}=10$ binning, estimated by PRIMA ETC}
\end{table} 

\section{Simulating Observable PAH Spectrum}
\label{sec:method}
We create a suite of simulated spectra of the PAH and dust grains heated by an interstellar radiation field and compare the observed flux density to the sensitivity of the PRIMA/FIRESS with low-resolution mode. In this section, we describe the methods and assumptions in our simulation.

\subsection{Model PAH emissivity}
\label{sec:pah_emissi}
A recent study by Draine et al., 2021\cite{draine_etal_2021} calculates the dust emission including PAHs for a range of the physical parameters of ISM and dust grains to provide the models for interpreting PAH spectra measured by ISO, Spitzer, AKARI, and JWST. We find that this library of PAH model spectra is also useful for simulating PAH spectra observed by PRIMA FIRESS. The spectrum of model PAH emissivity, $\nu P_{\nu}$ (erg s$^{-1}$ per H atom) is computed for a given set of parameters: spectral type of interstellar radiation field (ISRF), grain size distribution, PAH mass fraction ($q$PAH), PAH ionization fraction, scale parameter of ISRF strength ($U$) relative to the one in the solar neighborhood\cite{draine_etal_2021}. The total IR emissivity ($L_{\mbox{\tiny IR}}$) for dust and PAHs is also computed based on the choice of ISRF. For more detailed information, see Draine et al., 2021\cite{draine_etal_2021}.   

\begin{figure}
\begin{center}
\begin{tabular}{cc}
\includegraphics[width=0.50\textwidth]{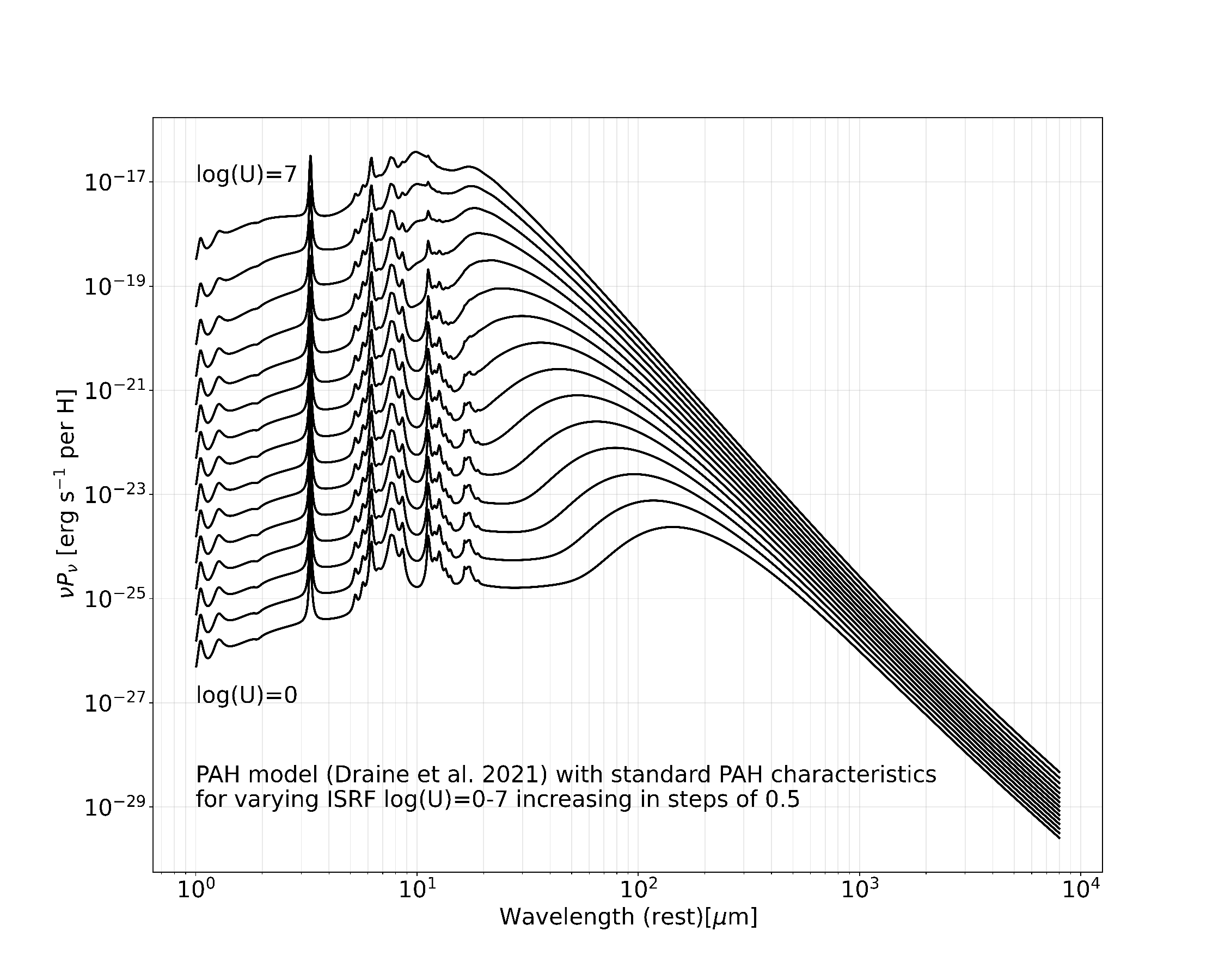} &
\includegraphics[width=0.48\textwidth]{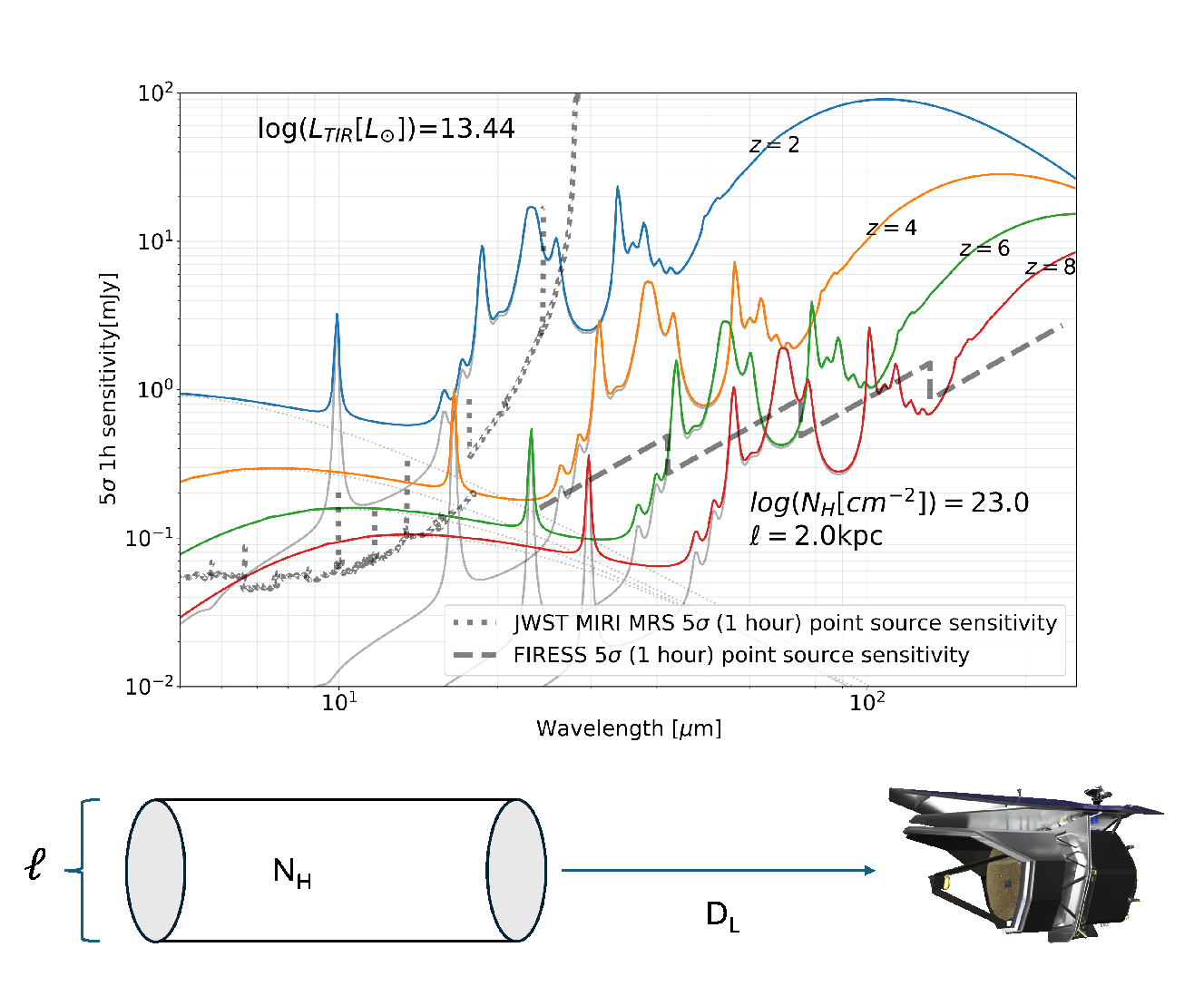}\\
\scriptsize (a) Emissivity of PAH and dust grains & \scriptsize (b) Simulation of observed spectra
\end{tabular}
\end{center}
\caption 
{\textit{Panel (a): }Model emissivity of the mixture of PAHs and dust grains per H atom\cite{draine_etal_2021} for `standard' grain size distribution, PAH ionization fraction, PAH mass fraction (qPAH). The emissivity is computed based on a source of heating by the modified version of the starlight spectrum in the solar neighborhood\cite{mathis_etal_1983} scaled up to $10^7$ times larger than the value in the solar neighborhood.  \textit{Panel (b): }Simulation of the observed spectrum from the model PAH and dust emissivity with $U=10^4$ (Panel (a)) for a model galaxy with $L_{\mbox{\tiny IR}}=10^{13.44} L_{\odot}$ and the chosen parameters of column density ($N_{\mbox{\tiny H}}=10^{23}$cm$^{-2}$), size of the emitting region ($\ell=2$kpc), and luminosity distance ($D_L$). The PAH emissivity uses mMMP ISRF, a standard grain size distribution, and PAH ionization fraction, and $q$PAH$=0.038$. The thick gray dotted and the thick gray dashed line indicate a $5\sigma$ point source sensitivity (with 1-hour integration) of JWST/MIRI and PRIMA/FIRESS, respectively.}\label{fig:pah_model_spec_illustration}
\end{figure}

Fig.~\ref{fig:pah_model_spec_illustration}(a) presents the emissivity spectra of PAH and dust grains\cite{draine_etal_2021}, $\nu P_{\nu}$ for a range of ISRF strength $U$ ($1<U<10^7$) for the ISRF spectrum in the solar neighborhood (mMMP\cite{mathis_etal_1983}), `standard'\footnote{The term `standard' means the standard parameter values for grain size distribution function listed in Table 2 in Draine et al., 2021\cite{draine_etal_2021}.} grain size distribution, PAH mass fraction ($q$PAH=$0.038$) and PAH ionization fraction.  The continuum spectrum of ISRF corresponding to each $U$ value as a heating source of PAH and dust grains is not shown in Fig.~\ref{fig:pah_model_spec_illustration}(a) but included in the simulation of the observed spectrum (see Section~\ref{sec:pah_spec_illust}). The PAH features and the continuum spectral shape are affected (i.e., weaker PAH emission and longer FIR peak wavelength) if one considers the attenuation of starlight incident on dust clouds\cite{draine_etal_2021}. For mMMP ISRF (used in this study) illuminating a foreground dust slab with A$_V=2.0$, the PAH features are weakened by a factor of $\sim1.5$ for $U=10^3$ while the impact of attenuation on the PAH emission is negligible for the 3 Myr old `young' starburst ISRF where the most of the stellar power is in the far-UV\cite{draine_etal_2021}. The difference in the PAH features between the attenuated and the unattenuated starlight is likely to be smaller for stronger ISRF (i.e., larger $U$) since the starlight from stronger ISRF can penetrate deeper. Since the model\cite{draine_etal_2021} computes the attenuated PAH spectra with a foreground dust slab for only one A$_V$ value (A$_V=2.0$), we cannot create the model PAH spectra for the ISRF with different dust attenuations. Also, the average strength of mMMP ISRF considered in the analysis (Section~\ref{sec:result}) is $U\sim10^{3.5}$. Therefore, we expect that the effect of dust attenuation on the impinging starlight is small and adopt the model PAH emissivity spectra from the unattenuated ISRF.   

\subsection{PRIMA FIRESS Sensitivity}
\label{sec:firess_sensi}
The instrument parameters of the FIRESS base grating module\cite{bradford_etal_2024} are given in Table~\ref{tab:instpar}. The nominal value of the point source sensitivity for an unresolved line (i.e., line sensitivity) with $5\sigma$ in 1-hour integration is $1.9\times10^{-19}$ W m$^{-2}$. The nominal $5\sigma$ point source continuum sensitivities in four observing bands for $R=10$ binning in 1-hour integration, estimated from the PRIMA Exposure Time Calculator\footnote{\linkable{https://prima.ipac.caltech.edu/page/etc-calc}}, are also listed in Table~\ref{tab:instpar}. Using the spectral sampling (i.e., resolution) parameter for each band, we estimate the spectral sensitivity in units of Jy ($10^{-26}$W m$^{-2}$ Hz$^{-1}$ or $10^{-23}$erg s$^{-1}$ cm$^{-2}$ Hz$^{-1}$) as shown by the thick gray dashed line in Fig.~\ref{fig:pah_model_spec_illustration}(b). 

Fig.~\ref{fig:pah_model_spec_illustration}(b) illustrates the observed mid-infrared (MIR) spectra of a galaxy at different redshifts by colored solid lines (see Section~\ref{sec:pah_spec_illust} for detailed prescription for computing the spectrum). For comparison, in Fig.~\ref{fig:pah_model_spec_illustration}(b), we also present a $5\sigma$ point source sensitivity with 1-hour integration for JWST/MIRI MRS (thick gray dotted line) scaled from the $10\sigma$ JWST/MIRI MRS sensitivity with 10,000 sec integration\footnote{\linkable{https://jwst-docs.stsci.edu/jwst-mid-infrared-instrument/miri-performance/miri-sensitivity}}. The sensitivity of the MIRI MRS becomes poor with increasing wavelength and makes the use of Channel 4 ($>18\mu$m) very difficult for observing high-redshift PAH emission. In contrast, the PRIMA/FIRESS sensitivity (thick gray dashed line) for a point source is 1-2 orders of magnitude better than that of the long-wavelength channel of JWST/MIRI MRS and extends the redshift limit for observing PAH emissions to $z>4$.

\subsection{Observed PAH spectrum}
\label{sec:pah_spec_illust}
The computation of the observed flux density at redshift $z$ from the emissivity, $\nu P_{\nu}$ (erg s$^{-1}$ per H atom) requires gas column density ($N_{\mbox{\tiny H}}$[cm$^{-2}$]) and diameter of the emitting region ($\ell$) as illustrated in Fig.~\ref{fig:pah_model_spec_illustration}(b). The spectral flux density, $F_\nu$ (Jy) is computed as 
\begin{equation}\label{eq:pahspec}
F_\nu = \frac{\pi}{4}\ell^2 N_{\mbox{\tiny H}} \frac{4\pi j_\nu}{n_{\mbox{\tiny H}}}\frac{1}{4\pi D_L^2}
\end{equation}where $j_\nu=\frac{1}{4\pi}P_\nu$ for isotropic emitter\cite{rybicki_lightman_1979}, $n_{\mbox{\tiny H}}$ is a volume density of hydrogen atom (H atom), and $D_L$ is a luminosity distance at redshift $z$ for a given cosmological model\footnote{In this study, we use the standard LCDM cosmological model with $H_0=70$ km s$^{-1}$ Mpc$^{-1}$, $\Omega_m=0.3$, $\Omega_\Lambda=0.7$.}. Although the line-of-sight dust extinction at V-band (A$_V$) is directly proportional to the gas column density ($N_{\mbox{\tiny H}}$), the dust attenuation due to a complex geometry of dust and stellar population is more complicated than the extinction and depends weakly on $N_{\mbox{\tiny H}}$ \cite{salim_narayanan_2020}. Also, we use the model spectra from unattenuated ISRF (Section~\ref{sec:pah_emissi}). Therefore, we do not link the $N_{\mbox{\tiny H}}$ to the attenuation of ISRF.

The term $\frac{4\pi j_\nu}{n_{\mbox{\tiny H}}}$ is given from the PAH-dust grain model\cite{draine_etal_2021} for the chosen parameters. In this study, we fix certain parameters: we use a modified version of the starlight spectrum in the solar neighborhood\cite{mathis_etal_1983} called mMMP\cite{draine_etal_2021}, `standard' grain size distribution and `standard' PAH ionization fraction for computation (see Figure 9(a) and (b) in Draine et al., 2021\cite{draine_etal_2021}). As discussed in Section~\ref{sec:pah_spec_dist} and \ref{sec:pah_spec_qpah}, we adopt different values of the scale parameter of the strength of ISRF ($U$) relative to the value in the solar neighborhood\cite{mathis_etal_1983} for a given column density and consider three different values of the mass fraction (relative to the total dust mass) of PAH molecules whose number of carbon atom is less than 1000, i.e., the parameter $q$PAH\cite{draine_li_2007}.

Since the PAH model spectra\cite{draine_etal_2021} do not extend to the rest-frame optical and near-infrared (NIR) wavelengths, we combine the chosen ISRF energy density spectrum\cite{draine_etal_2021} (mMMP in this study) scaled by $U$ and the PAH spectrum to construct the total spectrum. The total energy density spectrum of ISRF (erg cm$^{-3}$), $\nu u_\nu$ is converted to a luminosity (erg s$^{-1}$) by integrating over the volume of a spherical shell\footnote{Although the PAH spectrum is calculated based on a cylindrical geometry, a spherical geometry is used to calculate ISRF spectrum because a volume of spherical shell based on spherical geometry captures the ISRF distribution better than a volume of flat disk based on cylindrical geometry. For the same radius, the volume of a spherical shell is four times larger than that of a flat disk.} with a radius $\ell/2$ and a thickness $c$ as light propagation distance per unit time and the luminosity is converted to the observed flux density for a given luminosity distance $D_L$ at redshift $z$.  

As an illustration, Fig.~\ref{fig:pah_model_spec_illustration}(b) shows the model spectrum of the dust and PAH molecules, including the ISRF spectrum for a galaxy observed at four different redshifts ($z=2, 4, 6, 8$) represented by the solid lines with different colors. For each colored solid line, the thin gray dotted lines show ISRF spectra, and the thin solid gray lines show the PAH-dust spectra. Each spectrum is created using the same parameter value ($N_{\mbox{\tiny H}}=10^{23}$cm$^{-2}$, $\ell=2$kpc, and $U=10^4$). The total IR luminosity of the galaxy is obtained by scaling the total IR power (erg s$^{-1}$ per H atom) of each model spectrum model using the ISRF scale parameter $U$ and converting it to the luminosity.

\begin{figure}
\begin{center}
\begin{tabular}{cc}
\includegraphics[width=0.45\textwidth]{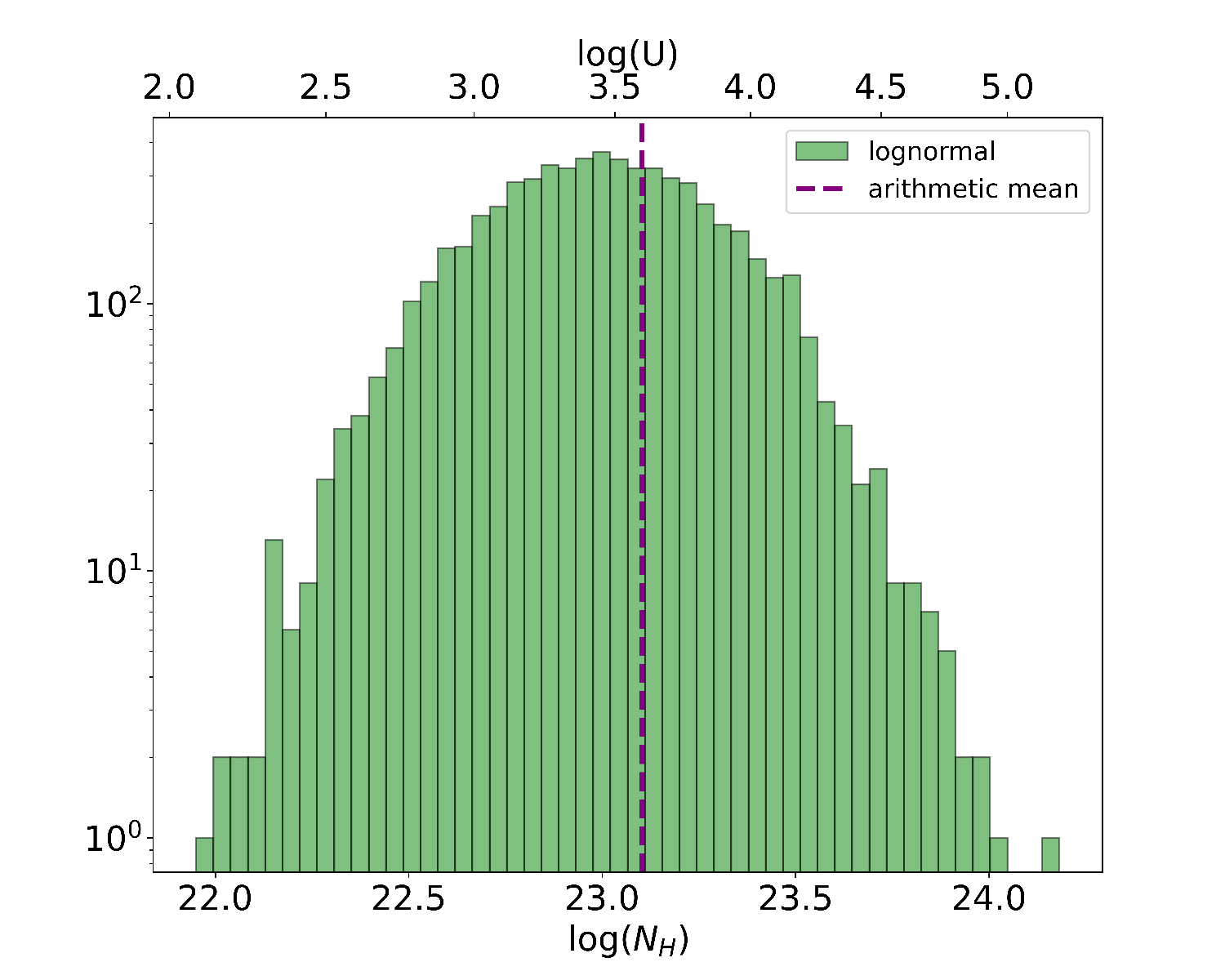} &
\includegraphics[width=0.55\textwidth]{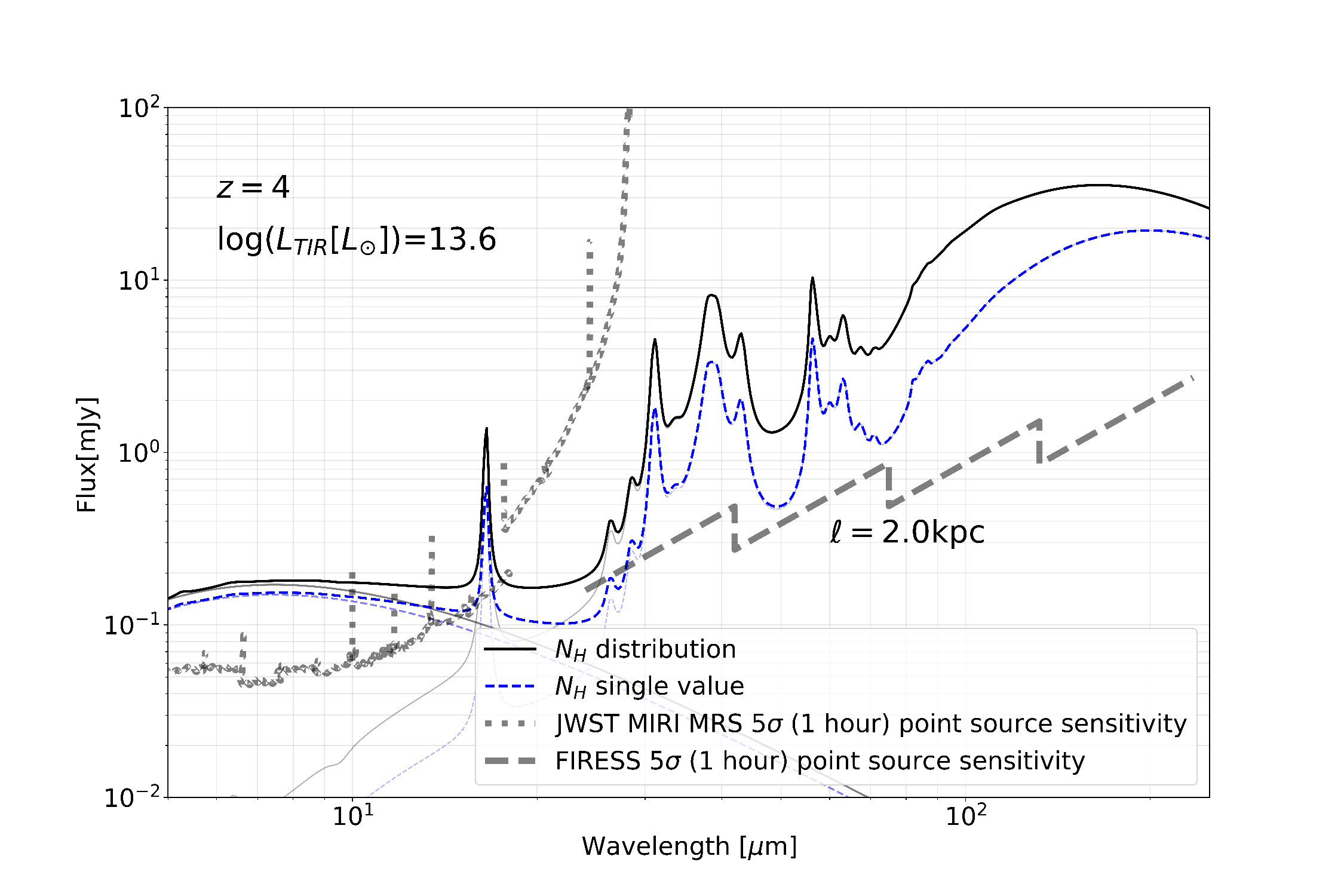}\\
\scriptsize (a) Column density distribution  & \scriptsize (b) Model spectra at $z=4$
\end{tabular}
\end{center}
\caption 
{\textit{Panel (a): }Log-normal distribution of column density $N_{\mbox{\tiny H}}$ with a variance $\sigma_s=0.7$, sampled by 6000 random numbers centered at log$(N_{\mbox{\tiny H}}[\mbox{cm}^{-2}])=23$. The purple vertical line indicates the arithmetic mean of $N_{\mbox{\tiny H}}$ values.  \textit{Panel (b): } Model PAH spectrum of a galaxy at $z=4$ with $q$PAH=0.038. The black solid line is obtained by ensemble averaging of the individual spectrum with the chosen $N_{\mbox{\tiny H}}$ and associated $U$ while the spectrum with the blue dashed line is obtained by the representative (i.e., arithmetic mean) value of $N_{\mbox{\tiny H}}$ and $U$ from the distribution.
}\label{fig:spec_dist_lognorm}
\end{figure} 

\begin{figure}
\begin{center}
\begin{tabular}{cc}
\includegraphics[width=0.45\textwidth]{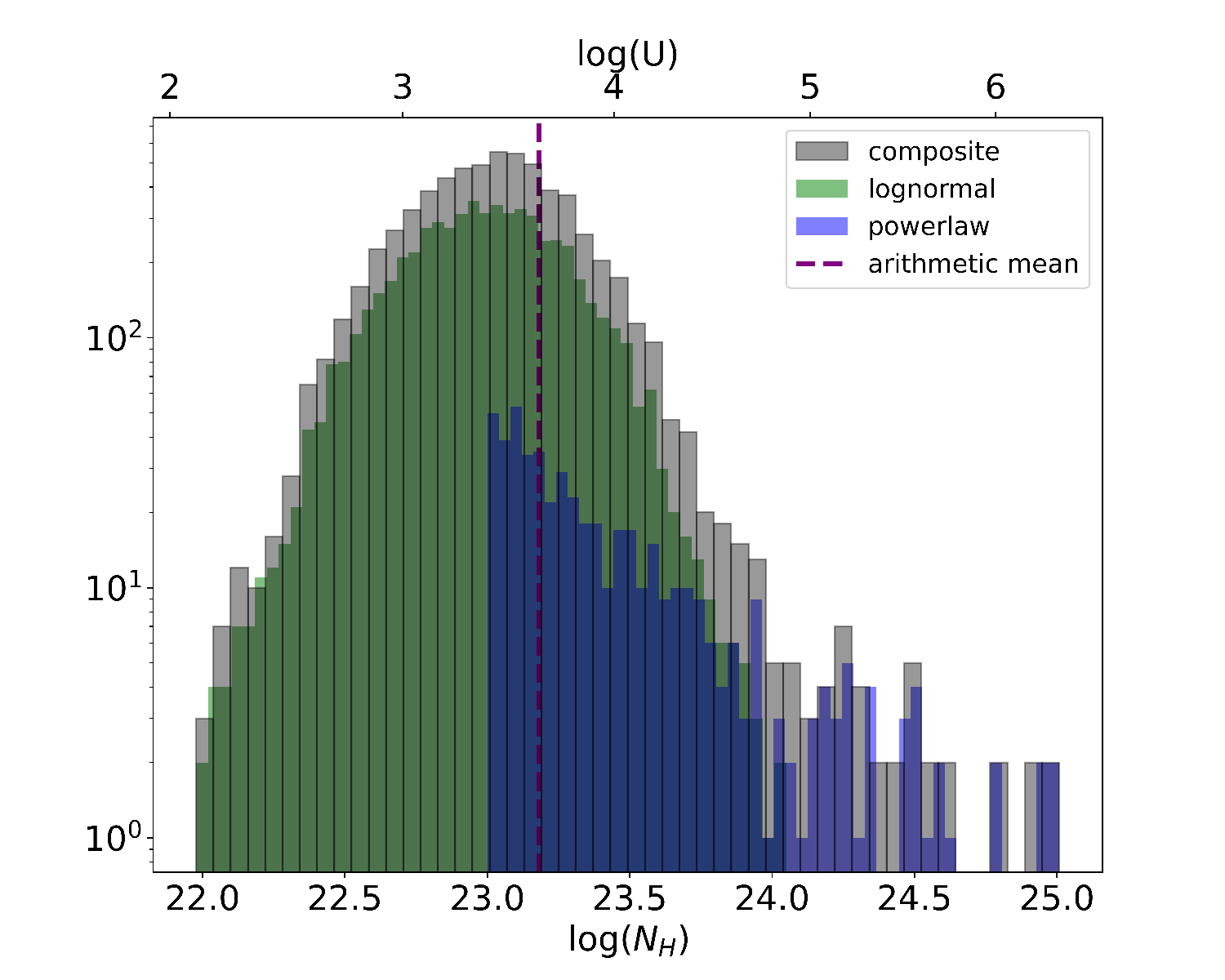} &
\includegraphics[width=0.55\textwidth]{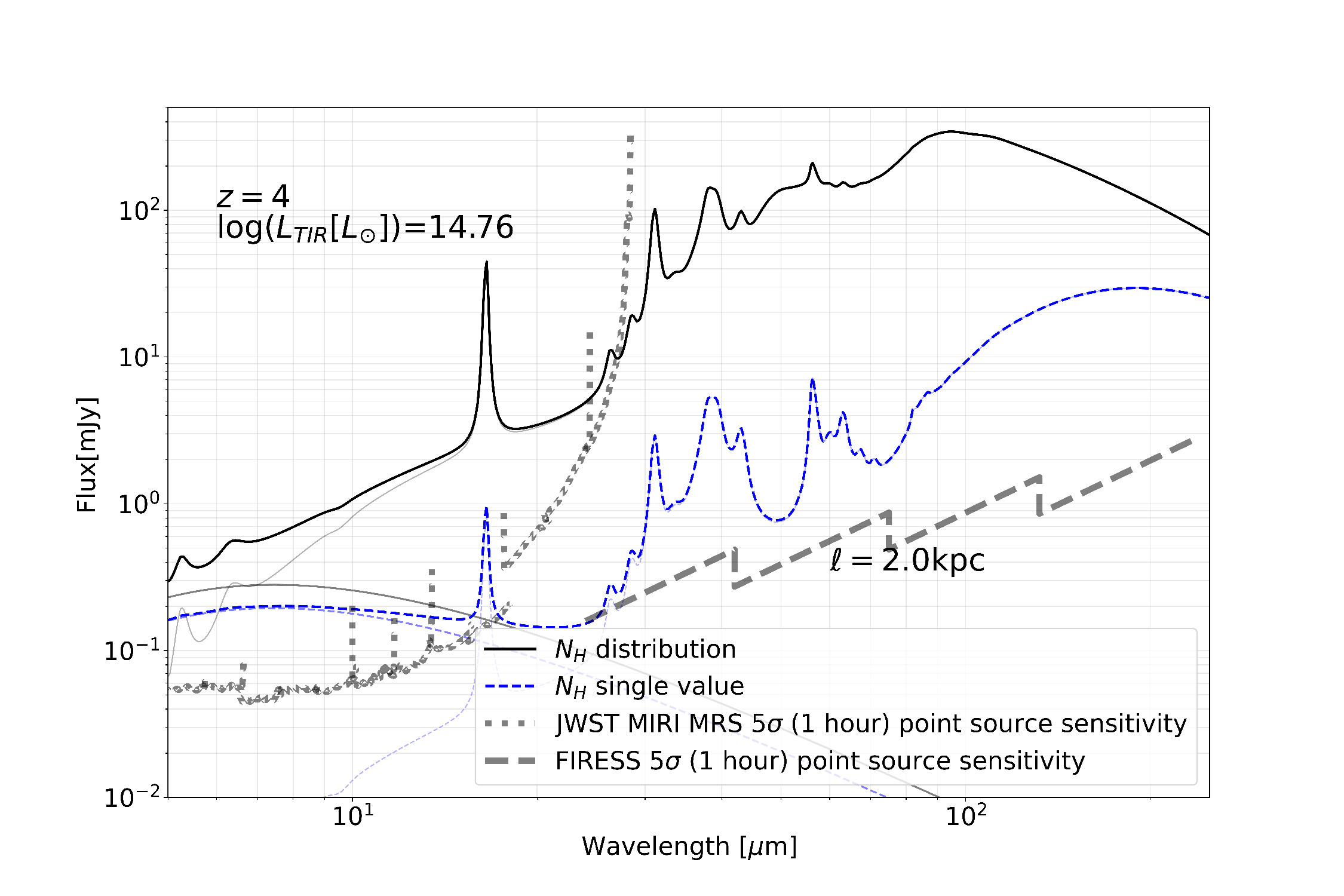}\\
\scriptsize (a) Column density distribution & \scriptsize (b) Model spectra at $z=4$
\end{tabular}
\end{center}
\caption 
{Same as Fig.~\ref{fig:spec_dist_lognorm}, but the column density distribution is a sum (shown by gray histogram) of log-normal (6000 random numbers shown by green histogram) and power-law distribution (using 500 random numbers shown by blue histogram) with a spectral slope of -2.5 .}\label{fig:spec_dist_logpow}
\end{figure} 

\subsection{Distributions of column density and ISRF}
\label{sec:pah_spec_dist}
PRIMA with a 1.8m diameter mirror does not spatially resolve a high-redshift galaxy. The pixel sizes of the PRIMA/FIRESS are 7.6\arcsec\ in Band 1/2, 12.7\arcsec\ in Band 3, and 22.9\arcsec\ in Band 4 (see Table~\ref{tab:instpar}), which implies that the observed PAH spectrum is a mixture of the contributions from many star-forming regions and diffuse ISM with different gas column densities and ISRF. Therefore, a PAH spectrum based on an ensemble of individual spectra from different regions is a more realistic model than a spectrum based on a single representative value for $N_{\mbox{\tiny H}}$ and $U$.  

It has been known that from observations and simulations, the column density probability distribution (N-PDF) is an important diagnostic of turbulence in local star-forming clouds\cite{federrath_etal_2010,burkhart_etal_2015b} and traces SFR\cite{burkhart_etal_2015a,schneider_etal_2015}. The two-component (lognormal and power law distribution) model of N-PDF is widely used to describe the observed N-PDF\cite{chen_etal_2018,kortgen_etal_2017}. A consensus is that the N-PDF of low column density gas and diffuse ISM follows a lognormal distribution while that of high column density gas and warm ISM follows a power law distribution\cite{chen_etal_2018}. 

The far-UV emission (FUV) is produced by OB associations so that the ISRF intensity will be proportional to the SFR surface density \cite{ostriker_etal_2010}. This relationship between ISRF and SFR surface density ($\Sigma_{\mbox{\tiny SFR}}$) can be refined and expressed as a function of inter-source opacity by considering a simple model of dust absorption in galactic disk\cite{bialy_2020}. If we use a tight correlation between $\Sigma_{\mbox{\tiny SFR}}$ and gas surface density ($\Sigma_{\mbox{\tiny gas}}$)\cite{kennicutt_1998} converted from $N_{\mbox{\tiny H}}$ ($N_{\mbox{\tiny H}}\approx10^{20}$cm$^{-2}$$\Sigma_{\mbox{\tiny gas}}$/(1 M$_{\odot}$pc$^{-2}$)\cite{rahmati_etal_2013}), we can parameterize the ISRF using $N_{\mbox{\tiny H}}$ and $\Sigma_{\mbox{\tiny SFR}}$ value at the solar neighborhood ($\Sigma_{\mbox{\tiny SFR,0}}$) as follows.
\begin{equation}\label{eq:isrf}
    U=\left[2.5\times10^{-4}\left(\frac{N_{\mbox{\tiny H}}}{10^{20}\mbox{cm}^{-2}}\right)^{1.4}\right]\times\frac{3}{\Sigma_{\mbox{\tiny SFR,0}}}
\end{equation} where the factor 3 is chosen to be the value of the ratio between ISRF and $\Sigma_{\mbox{\tiny SFR}}$ at the critical inter-source opacity\cite{bialy_2020}.    
We sample the distribution of $N_{\mbox{\tiny H}}$ using lognormal (Fig.~\ref{fig:spec_dist_lognorm}(a)) and lognormal with power law (Fig.~\ref{fig:spec_dist_logpow}(a)) distribution which also determines the ISRF scale factor $U$. Theoretical models of a turbulent ISM\cite{padoan_etal_1997,federrath_etal_2008} suggest that the variance of the lognormal distribution, $\sigma_s$, depends on the ISM Mach number ($\sigma^2_s=\mbox{ln}(1+b\mathcal{M}^2)$ where $\mathcal{M}$ is the ISM Mach number). For each random sample of $N_{\mbox{\tiny H}}$ and $U$, we create a spectrum as explained in Section~\ref{sec:pah_spec_illust} by interpolating the spectrum in a grid of the $U$ values. Then we average the ensemble of spectra created from each random sample of $N_{\mbox{\tiny H}}$ and $U$ with the assumption that the turbulent ISM only changes the distribution of $N_{\mbox{\tiny H}}$ in the ISM and does not cause an effect on the PAH properties like grain size and ionization fraction. Here we note that the correlation between $N_{\mbox{\tiny H}}$ and $U$ may break at high column density because $\Sigma_{\mbox{\tiny SFR}}$ reaches the Eddington limit (e.g., $\Sigma_{\mbox{\tiny SFR}}\approx3000$ M$_{\odot}$yr$^{-1}$kpc$^{-2}$ as a theoretical limit for starburst galaxies\cite{murray_etal_2005,thompson_etal_2005}) and disrupts the surrounding gas. We verify that the required $N_{\mbox{\tiny H}}$ to reach $\Sigma_{\mbox{\tiny SFR}}\approx3000$ M$_{\odot}$yr$^{-1}$kpc$^{-2}$ based on the KS relation\cite{kennicutt_1998}($N_{\mbox{\tiny H}}=1.48\times10^{25}\mbox{cm}^{-2}$) is lower than the maximum column density ($N_{\mbox{\tiny H}}=10^{25}\mbox{cm}^{-2}$) sampled from lognormal distribution with high density power law tail (Fig.~\ref{fig:spec_dist_logpow}(a)). Also, we do not use the PAH spectra created from the lognormal$+$power law density distribution in our analysis (Section~\ref{sec:result}). Therefore, the simulations in this study based on the simple scaling relation (Equation~\ref{eq:isrf}) are performed for a physically reasonable parameter range.

A model spectrum of a galaxy at $z=4$ computed based on a $N_{\mbox{\tiny H}}$ density distribution following a lognormal distribution centered at $N_{\mbox{\tiny H}}=10^{23}$cm$^{-2}$ with $\sigma_s=0.7$ (Fig.~\ref{fig:spec_dist_lognorm}(b)) has log($L_{\mbox{\tiny IR}}/L_{\odot}$)$=13.6$. This value is similar to the observed (without lensing magnification correction) $L_{\mbox{\tiny IR}}$ value\cite{debreuck_etal_2019}(log($L_{\mbox{\tiny IR}}/L_{\odot}$)$=13.9$) of a galaxy at $z=4.22$ of which PAH emission was observed (3440 sec observing time) by JWST/MIRI MRS\cite{spilker_etal_2023} and has a peak flux density (0.3 mJy) of the total (line and continuum) spectrum similar to the same model spectrum with lower $q$PAH value (0.5\%, see Fig.~\ref{fig:spec_qpah_z4}(a)). A model spectrum of the same galaxy with the same lognormal and an additional small fraction of power-law tail with a spectral slope of -2.5 (Fig.~\ref{fig:spec_dist_logpow}(b)) has much larger $L_{\mbox{\tiny IR}}$ value (log($L_{\mbox{\tiny IR}}/L_{\odot}$)$=14.76$) which is similar to the $L_{\mbox{\tiny IR}}$ value of the most IR luminous galaxy W2246-0526\cite{tsai_etal_2015} at $z=4.6$ (log($L_{\mbox{\tiny IR}}/L_{\odot}$)$=14.54$\cite{harrington_etal_2025}). The range of IR luminosity ($10^{12-15}$$L_{\odot}$) discussed in Section~\ref{sec:detection_redshift} and the $L_{\mbox{\tiny IR}}$ value ($10^{13.6}$$L_{\odot}$) used for simulating the spectra presented in Section~\ref{sec:pah_measurement} are not far from reality: for example, high-redshift lensed dusty star-forming galaxies (PASSAGES\cite{kamieneski_etal_2024}) have intrinsic $L_{\mbox{\tiny IR}}\gtrsim 10^{13}L_{\odot}$ after magnification correction (median $\mu=7$), implying much higher `apparent' brightness.

Remarkably, the spectrum with lognormal $N_{\mbox{\tiny H}}$ distribution and additional small (8\%) contribution from high-density regions following a power law $N_{\mbox{\tiny H}}$ distribution (Fig.~\ref{fig:spec_dist_logpow}(b)) is very different from the one with the same lognormal $N_{\mbox{\tiny H}}$ distribution only (Fig.~\ref{fig:spec_dist_lognorm}(b)). Although the fraction is small, if there are regions with high $\Sigma_{\mbox{\tiny SFR}}$ that are likely to have high $U$, the PAH emissivity for those high $ U$ regions (e.g., can be as large as $>10 \times$ the mean value as seen in Fig.~\ref{fig:spec_dist_logpow}(a)) is at least an order of magnitude larger (see Fig.~\ref{fig:pah_model_spec_illustration}(a)) and therefore contribution from the high $U$ regions to the average spectrum becomes significant. This is an advantage for observing high-redshift galaxies with high $\Sigma_{\mbox{\tiny SFR}}$. The PAH spectrum with higher $U$ also has a significantly different spectral shape from the one with lower $U$ seen at $\approx 10\mu$m silicate feature. Therefore a mixture of the spectra with low and high $U$ might provide more flexibility to explain the observed ratio between different PAH bands.  

\begin{figure}
\begin{center}
\begin{tabular}{ccc}
\includegraphics[width=0.33\textwidth]{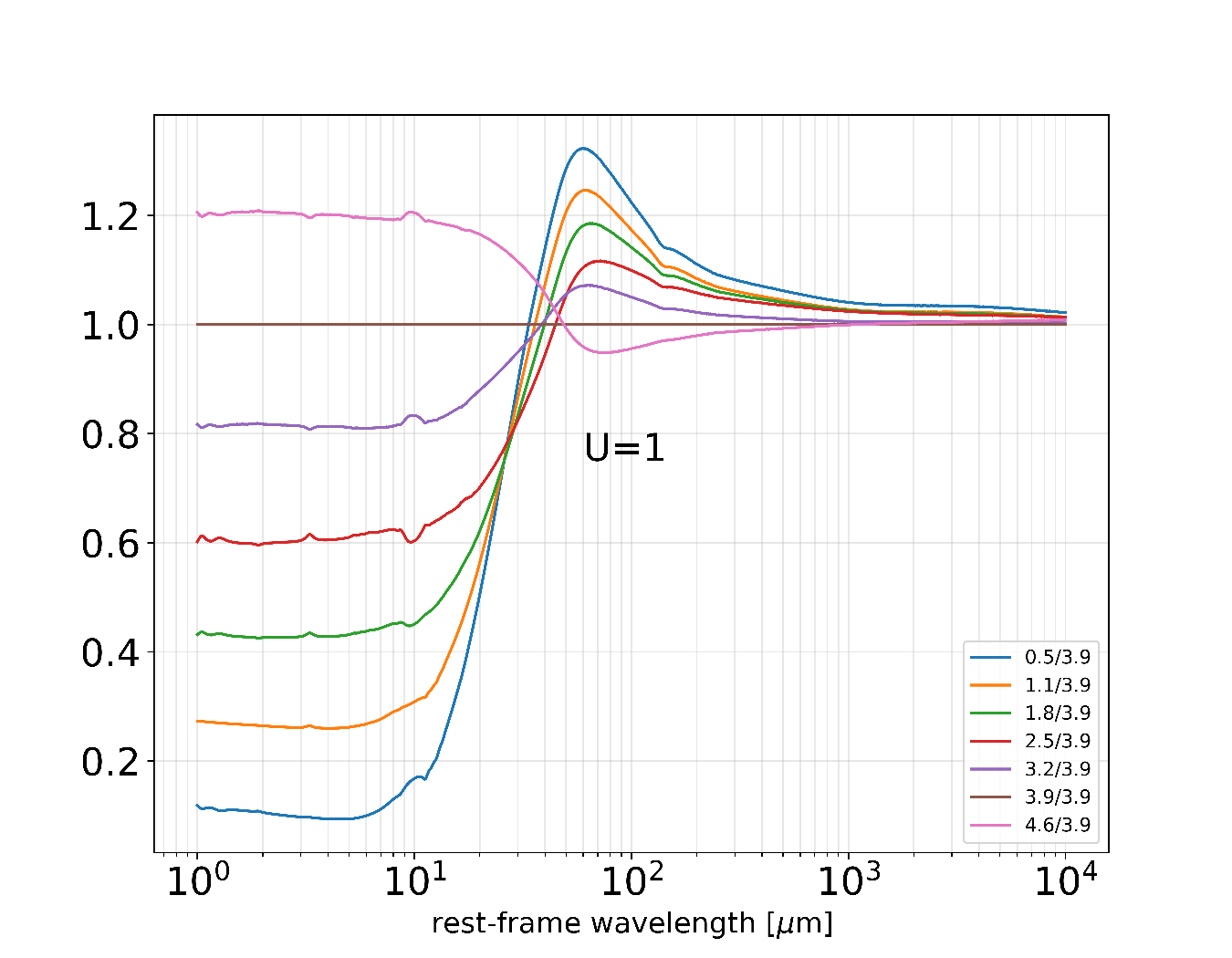} &
\includegraphics[width=0.33\textwidth]{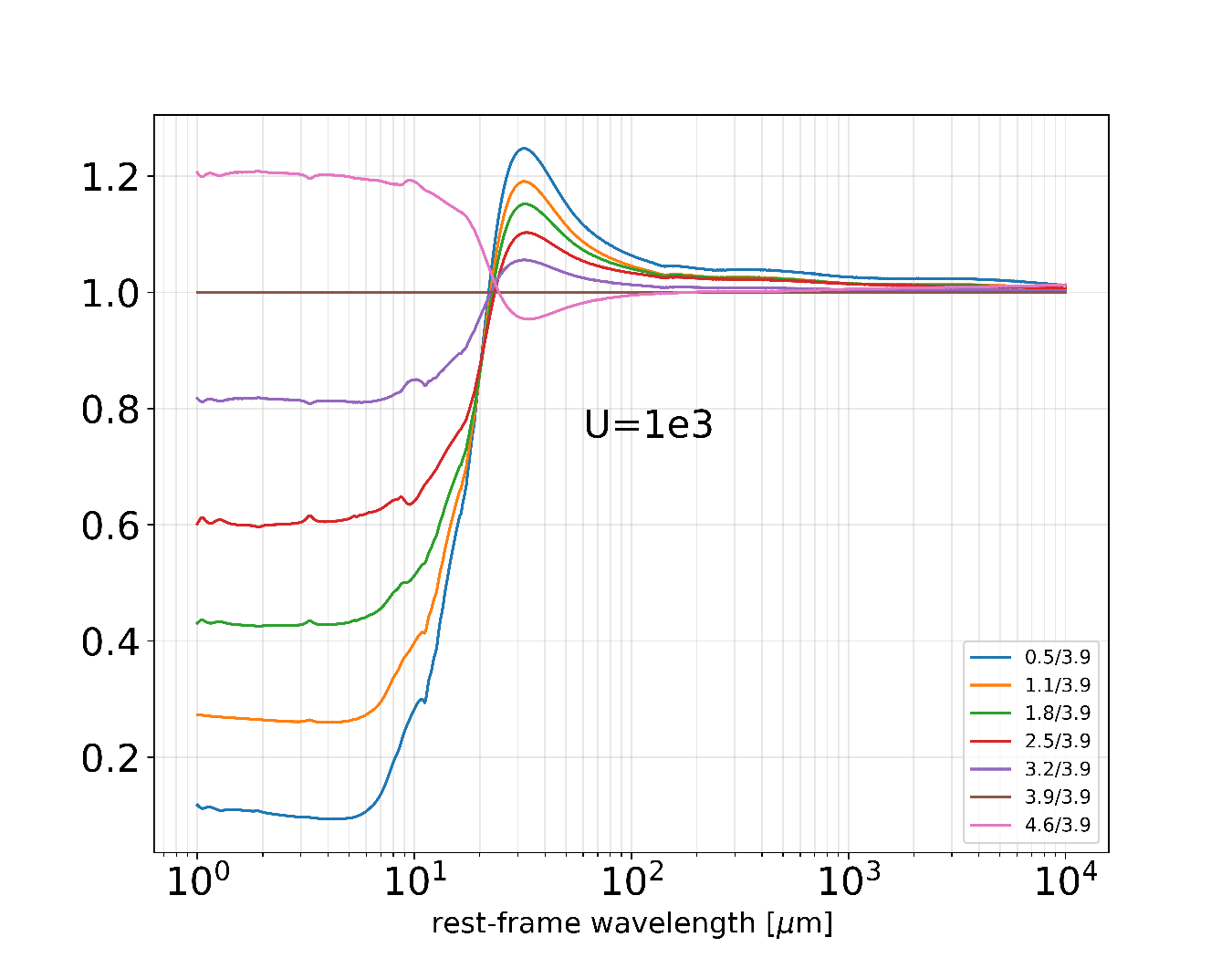} &
\includegraphics[width=0.33\textwidth]{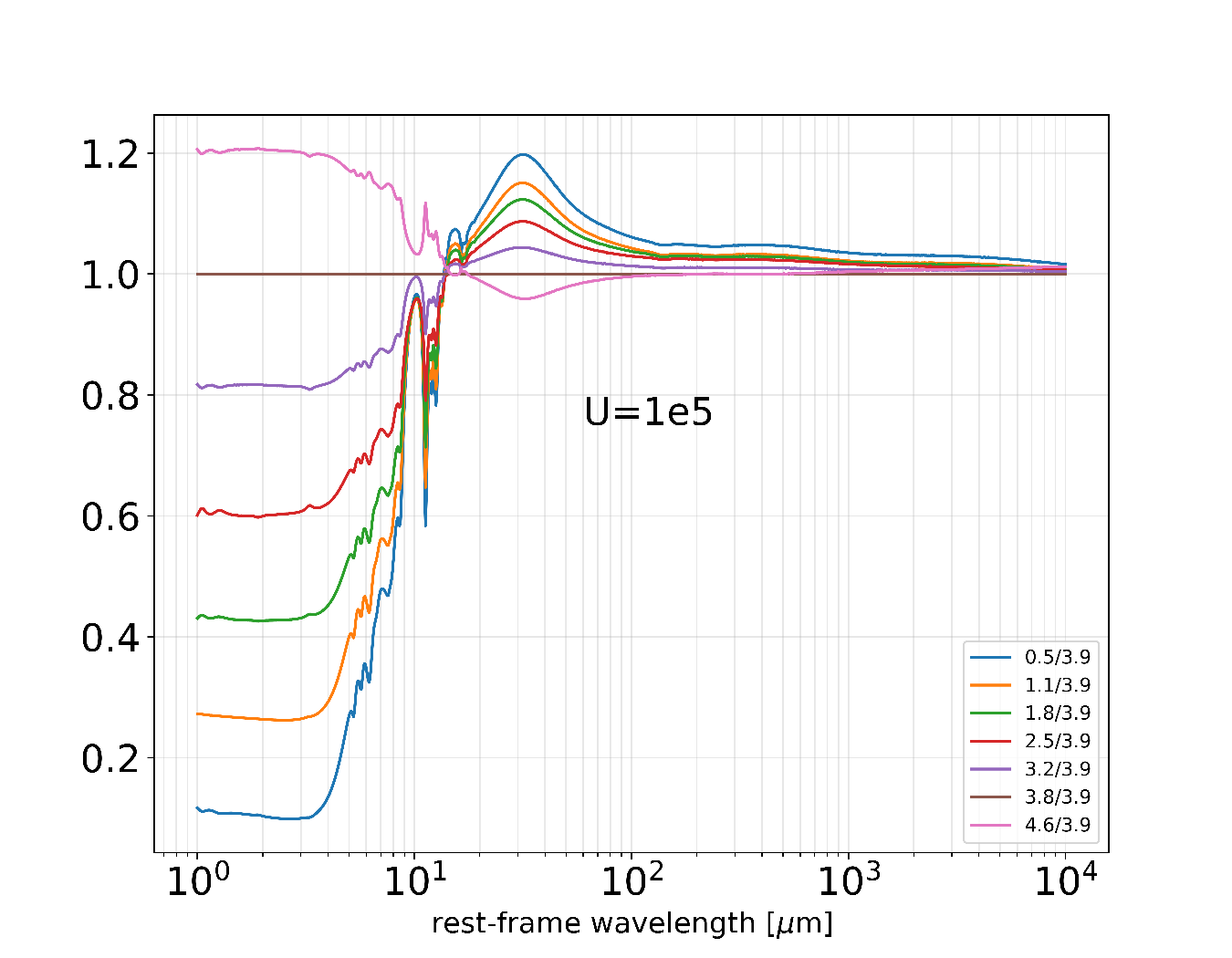} \\
\scriptsize (a) $U=1$ & \scriptsize (b) $U=10^3$ & \scriptsize (c) $U=10^5$ 
\end{tabular}
\end{center}
\caption 
{The PAH emissivity spectra with different $q$PAH values (color lines) normalized by the reference spectrum computed with $q$PAH$=0.039$ for different ISRF scale parameter $U$ using the model spectra\cite{draine_li_2007}}\label{fig:qpah_ratio}
\end{figure}

\subsection{Implementation of $q$PAH}
\label{sec:pah_spec_qpah}
A deficit of the fraction of PAH luminosity relative to total dust continuum luminosity is found at low-metallicity environments\cite{engelbracht_etal_2005,madden_etal_2006,draine_etal_2007,smith_etal_2007,munoz_etal_2009,sandstrom_etal_2012,aniano_etal_2020,shivaei_etal_2024,whitcomb_etal_2024}. This PAH-metallicity relation (PZR)\cite{whitcomb_etal_2024} implies that the inferred value of the mass fraction of PAH molecule with $<1000$ carbon atoms ($q$PAH parameter in the PAH model spectra\cite{draine_li_2007}) varies strongly with metallicity: lower $q$PAH with decreasing metallicity\cite{shivaei_etal_2024}. Although the correlation between $q$PAH and metallicity can be understood in a scenario where the hard radiation environment from high-mass stars in the low-metallicity ISM destroys preferentially the smallest grains\cite{madden_etal_2006,egorov_etal_2023}, the observed correlation between PAH emission and metallicity are also affected by the efficiency of grain growth and the resulting dust grain size distribution \cite{whitcomb_etal_2024}. 

Since galaxy gas metallicity has a clear trend with galaxy properties and redshift (lower mass galaxies at higher redshift tend to have lower metallicity\cite{maiolino_mannucci_2019}), we explore three different (low, intermediate, and high) $q$PAH values: 0.5, 1.8, 3.8\% of which corresponding metallicities from the PZR relation\cite{whitcomb_etal_2024} are $0.2, 0.5, 0.8 Z_{\odot}$ respectively. The PAH model in this study using standard grain size distribution\cite{draine_etal_2021} has the spectra computed for $q$PAH$=0.038$ only, which does not allow us to derive the model spectra with lower $q$PAH values. Thus we use the previous model\cite{draine_li_2007} that computes the PAH spectra with varying $q$PAH value for the nearly same ISRF (MMP\cite{mathis_etal_1983}) as the one in the current model and calculate the ratio between the spectrum with varying $q$PAH values relative to the reference spectrum with $q$PAH$=0.039$\footnote{We note that the model with varying $q$PAH\cite{draine_li_2007} value does not have the spectra with the exact $q$PAH (0.038) used in this study, however, the difference should be negligible.}. This ratio is applied to our model spectra (computed with $q$PAH$=0.038$ with standard grain size distribution).  

\begin{figure}
\begin{center}
\begin{tabular}{cc}
\includegraphics[width=0.49\textwidth]{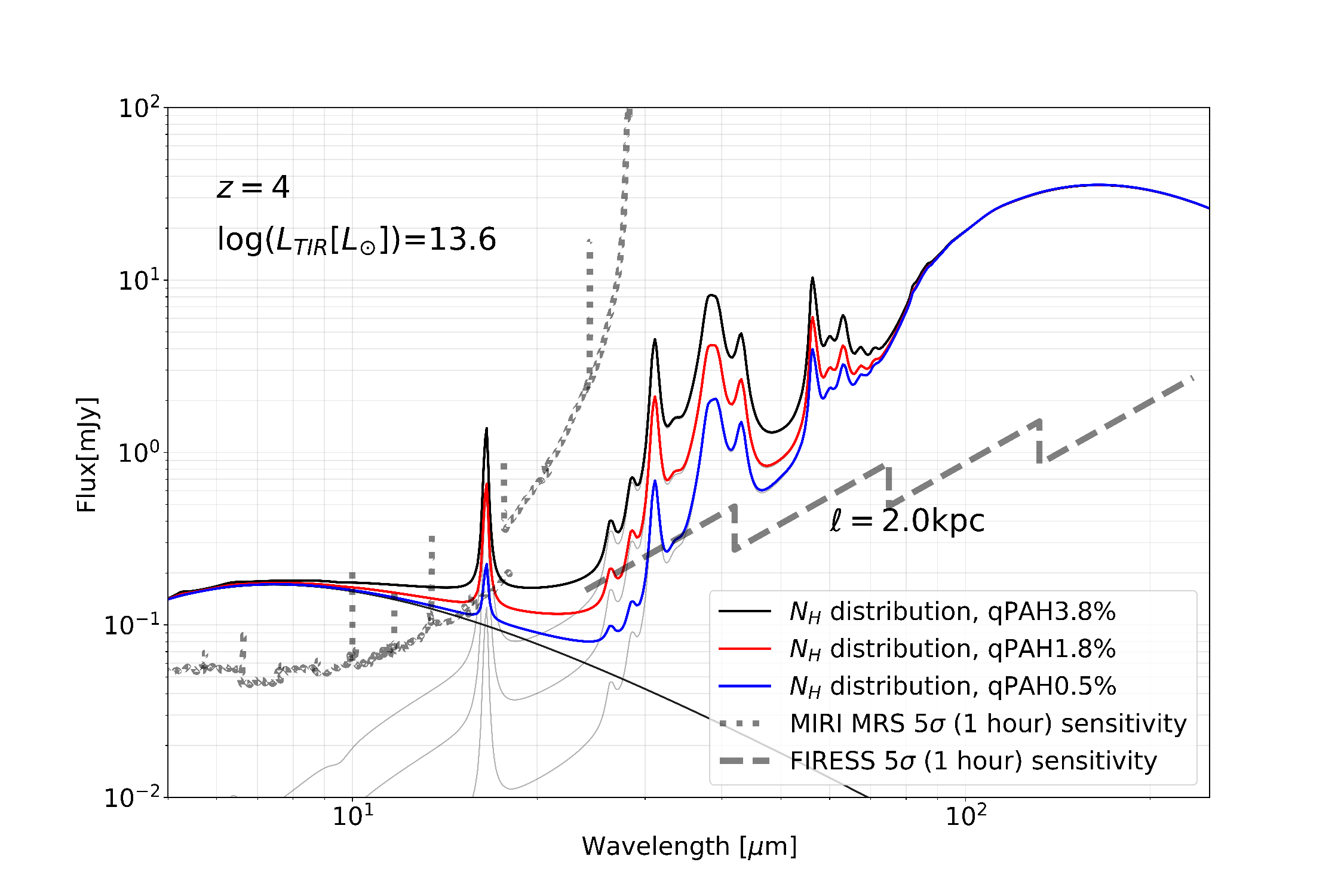} &
\includegraphics[width=0.49\textwidth]{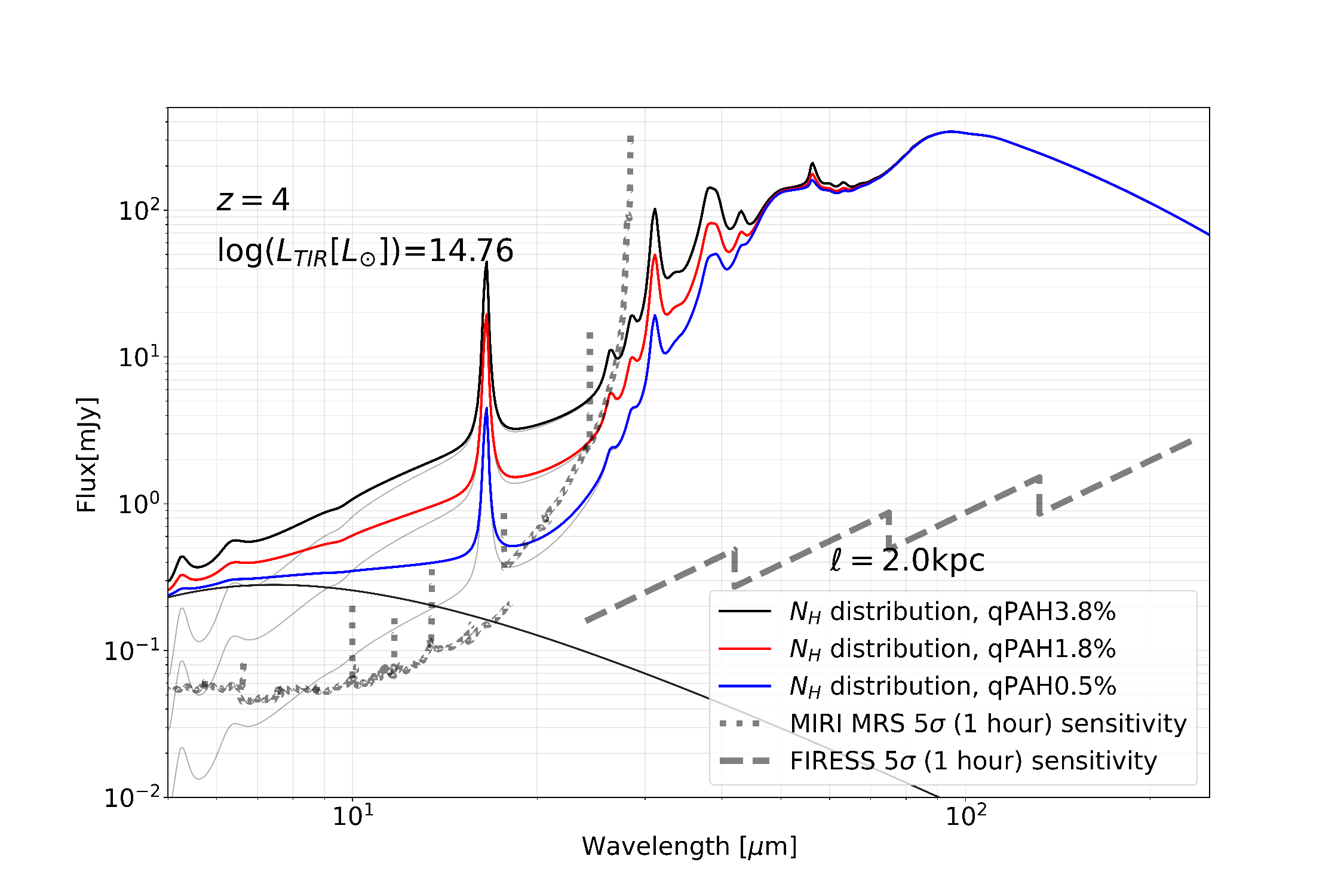}\\
\scriptsize (a) $N_{\mbox{\tiny H}}$ distribution in Fig~\ref{fig:spec_dist_lognorm} & \scriptsize (b) $N_{\mbox{\tiny H}}$ distribution in Fig~\ref{fig:spec_dist_logpow}
\end{tabular}
\end{center}
\caption 
{Model spectrum in Fig.3(b) and 4(b) using $N_{\mbox{\tiny H}}$ distribution (black solid line) and additional model spectra with the same parameters but different $q$PAH values: red line for 1.8\% and blue line for 0.5\%}\label{fig:spec_qpah_z4}
\end{figure}

Fig.~\ref{fig:qpah_ratio} presents the examples of PAH emissivity spectra with different $q$PAH values normalized by the reference spectrum with $q$PAH$=0.039$ computed using the model spectra\cite{draine_li_2007}. Three panels in Fig.~\ref{fig:qpah_ratio} show the case with low ($U=1$), intermediate ($U=10^3$), and high ($U=10^5$) $U$ parameters. The PAH emissivity spectra normalized by the reference spectrum with $q$PAH$=0.039$ have different shapes for different $U$ values. For simplicity, we do not assign $q$PAH for each sampled spectrum with $N_{\mbox{\tiny H}}$ value and globally apply the same $q$PAH value for the entire region of the galaxy. For a given $N_{\mbox{\tiny H}}$ sample distribution with the corresponding average $U$ value, we estimate the normalized PAH emissivity spectra to the reference spectrum (like the ones in Fig.~\ref{fig:qpah_ratio}) by interpolating the two normalized PAH emissivity spectra of which $U$ values from the model grid are bracketing the average $U$ value from the given $N_{\mbox{\tiny H}}$ distribution. Then the normalization of the PAH emissivity for two chosen $q$PAH values (0.005 and 0.018 in Fig.~\ref{fig:qpah_ratio}) are applied to the model spectrum computed with $q$PAH$=0.038$ to create two additional model spectra for low and intermediate $q$PAH values.  

Fig.~\ref{fig:spec_qpah_z4} shows the model spectra with three different $q$PAH values ($q$PAH$=0.005, 0.018, 0.038$). The model spectrum with $q$PAH$=0.038$ (black solid line) in Panel (a) and (b) are the same model spectrum in Fig.~\ref{fig:spec_dist_lognorm} and Fig.~\ref{fig:spec_dist_logpow}, respectively. In Panel (a) and (b), two additional spectra with low and intermediate $q$PAH value (blue and red solid lines) are shown. As expected, the resulting spectrum with lower $q$PAH value has weaker PAH emission in both line and continuum.  

\subsection{Simulation of the PRIMA/FIRESS spectrum with noise realization}
\label{sec:simul_firess_spec}
\begin{figure}
\begin{center}
\begin{tabular}{cc}
\includegraphics[width=0.50\textwidth]{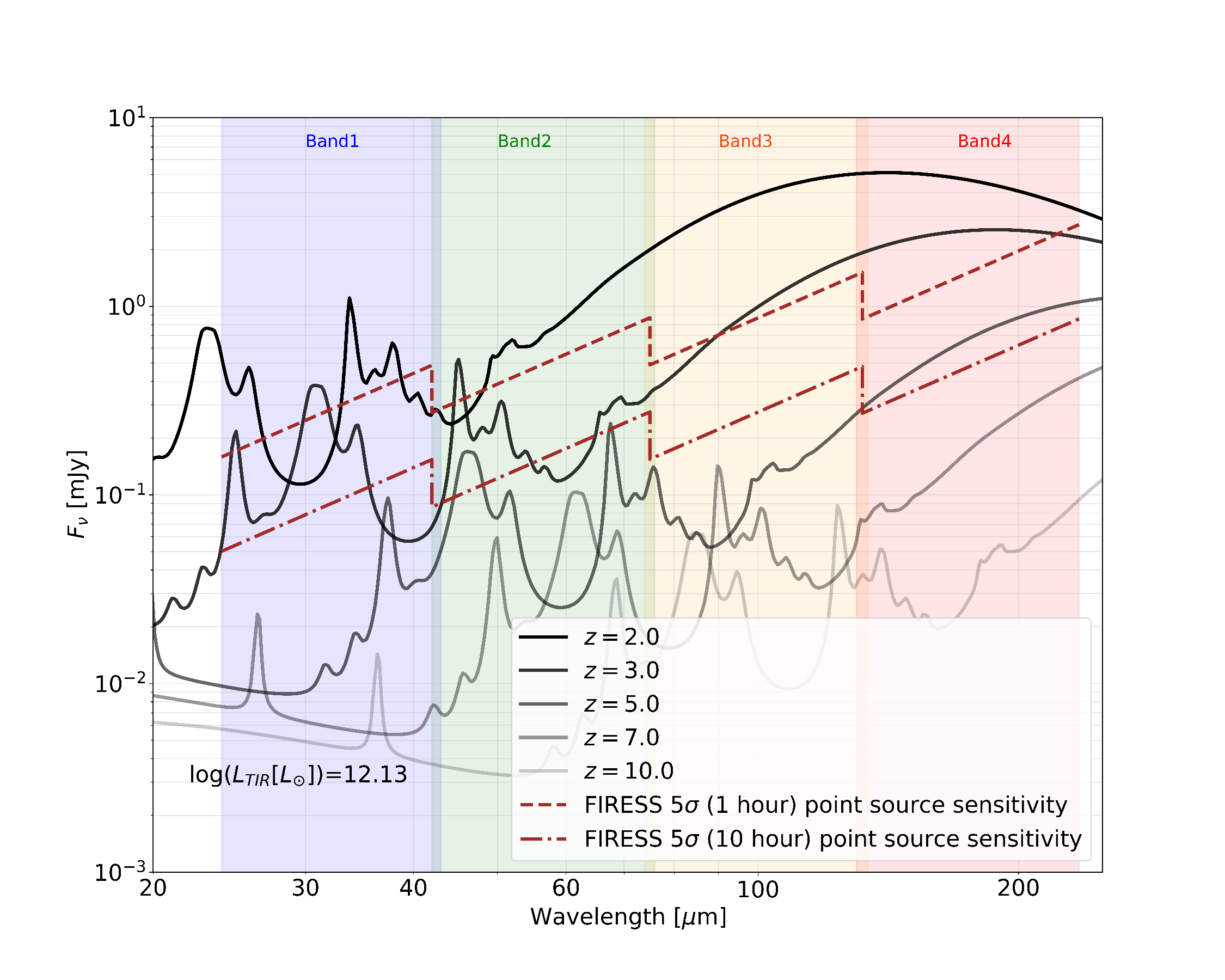} &
\includegraphics[width=0.50\textwidth]{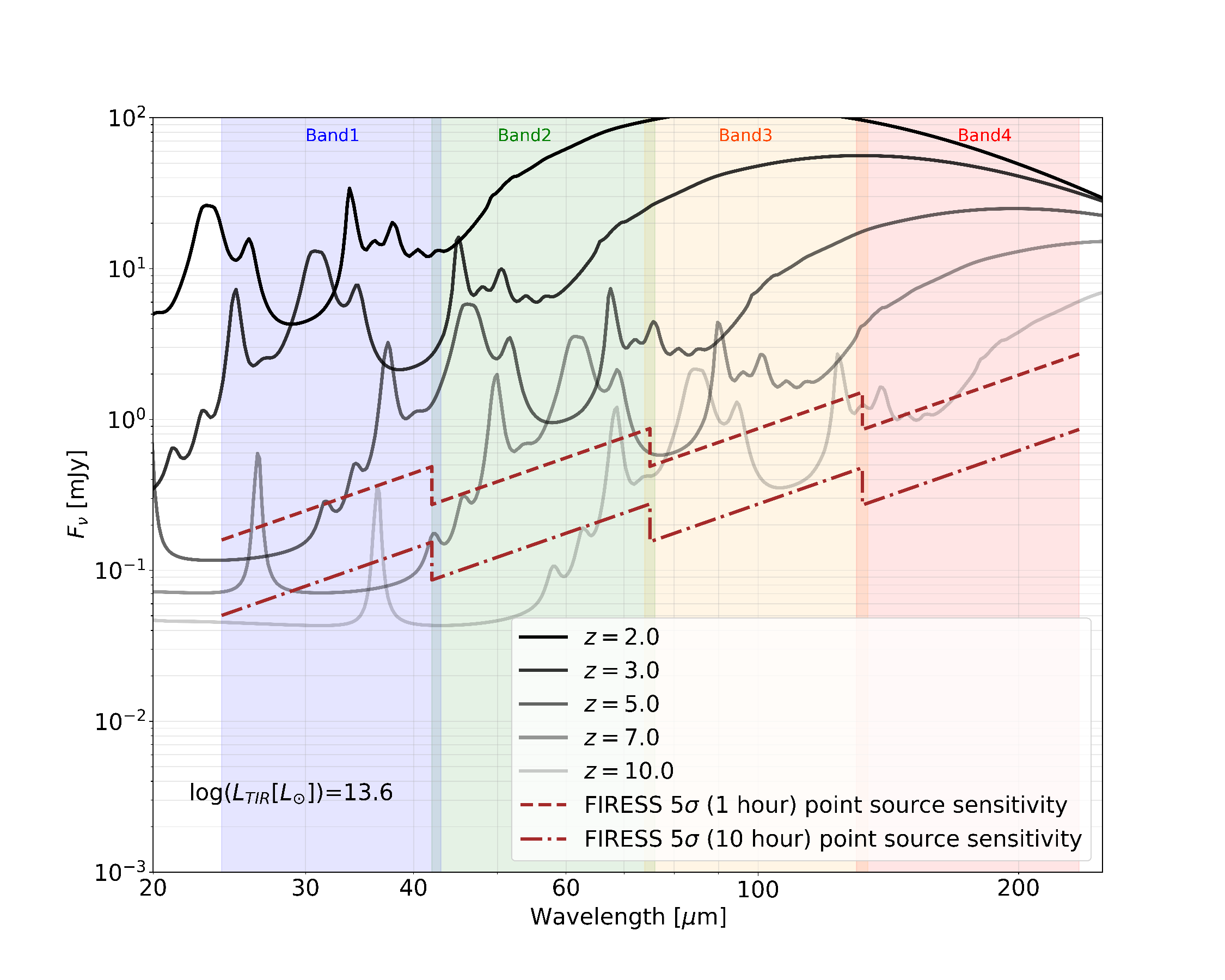} \\
\scriptsize (a) lognormal: $\bar{N}_{\mbox{\tiny H}}=2.5\times10^{22}$cm$^{-2}$ & \scriptsize (b) lognormal: $\bar{N}_{\mbox{\tiny H}}=1\times10^{23}$cm$^{-2}$ \\
\end{tabular}
\end{center}
\caption 
{Model spectrum of the PAH bands in the PRIMA/FIRESS observing bands at $z=2, 3, 5, 7, 10$. The spectrum is computed from three different lognormal distributions with the same variance ($\sigma_s=0.7$) and different means. $q$PAH$=0.038$ is used for all spectra. The purple dashed line and the purple dot-dashed line indicate a 5$\sigma$ sensitivity of the PRIMA/FIRESS from 1-hour and 10-hour integration, respectively.}\label{fig:firess_redshift}
\end{figure}

\begin{figure}
\begin{center}
\begin{tabular}{cc}
\includegraphics[width=0.49\textwidth]{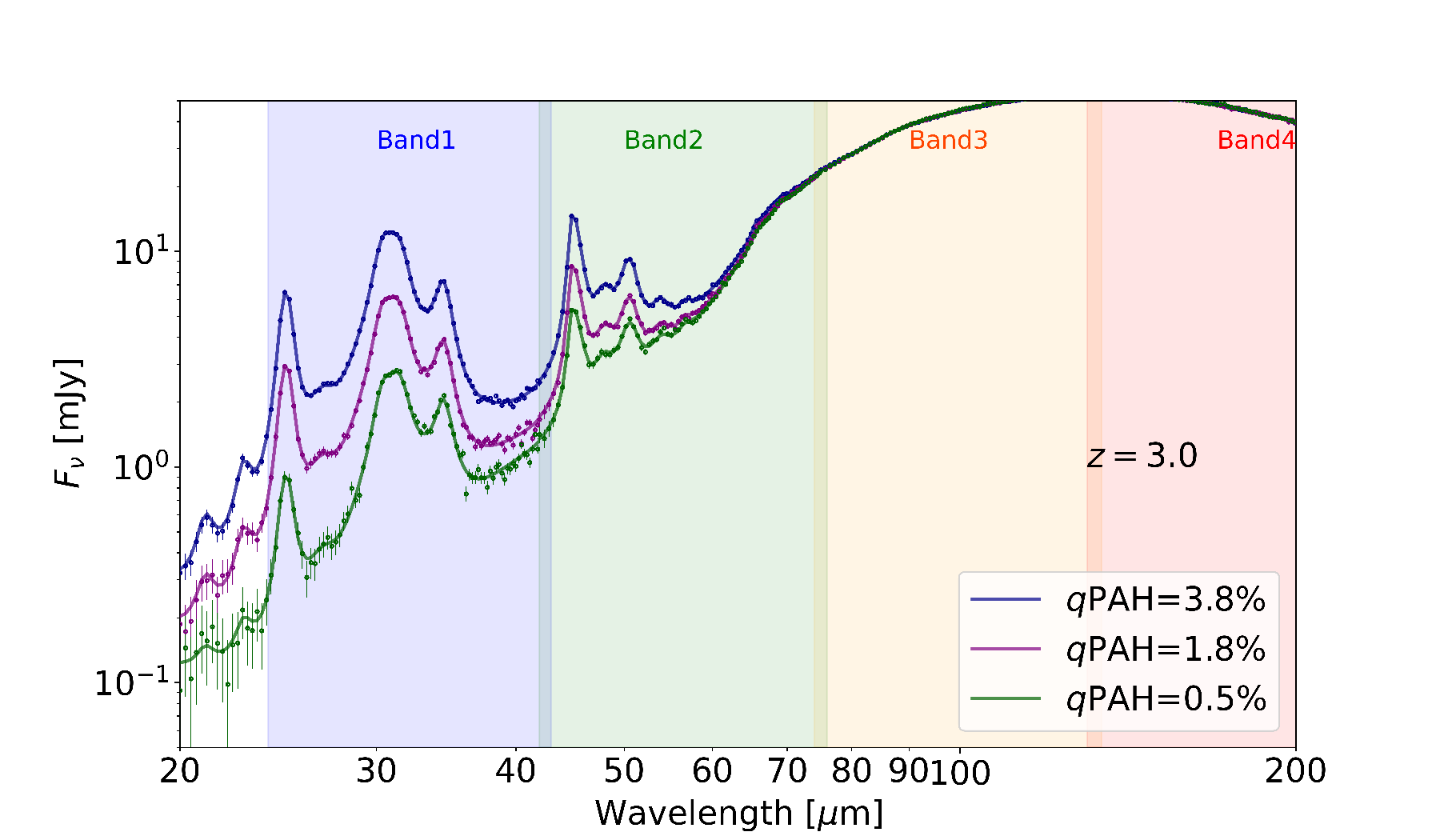} &
\includegraphics[width=0.49\textwidth]{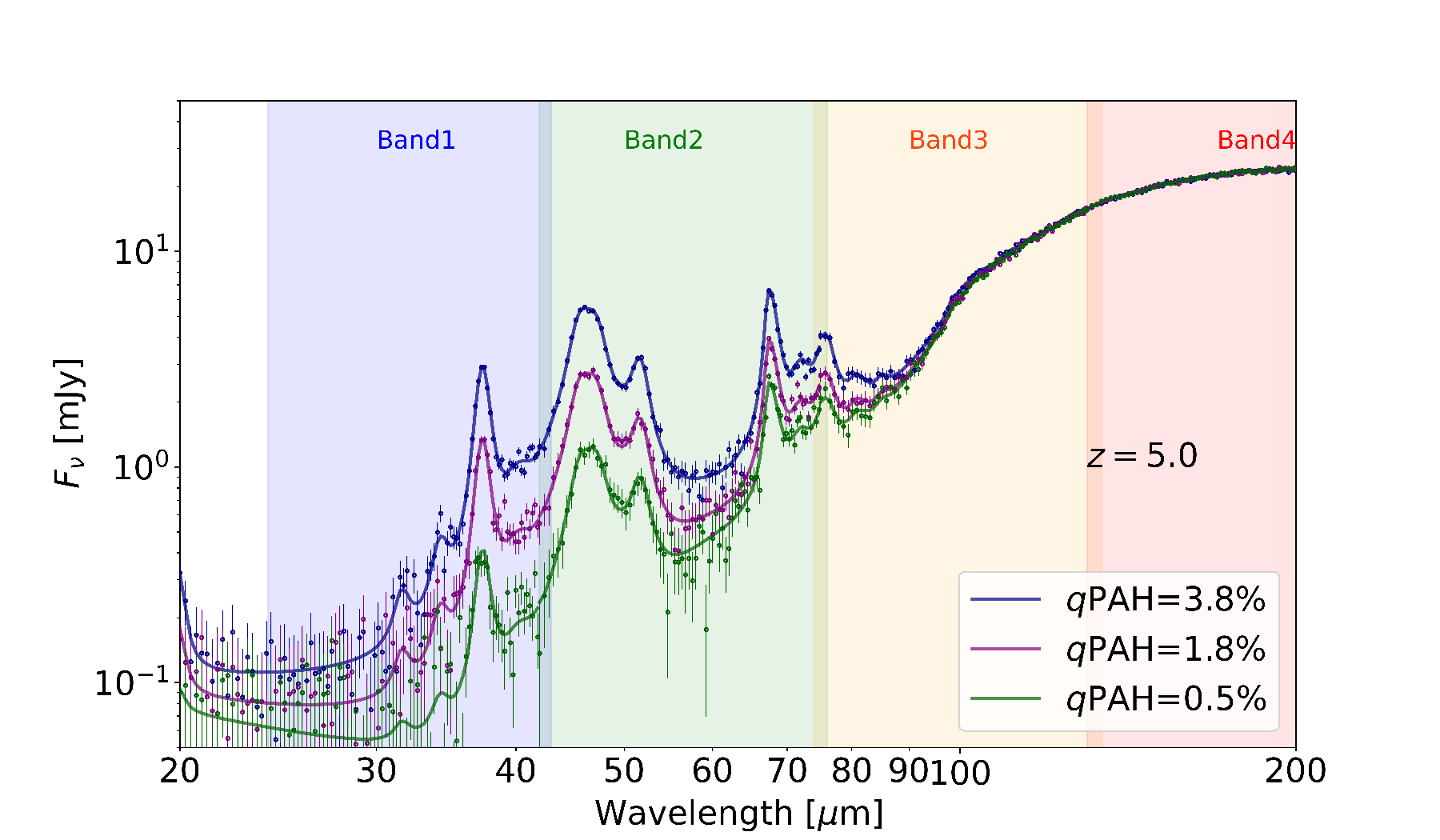}\\
\scriptsize (a) $z=3.0$ & \scriptsize (b) $z=5.0$ \\
\includegraphics[width=0.49\textwidth]{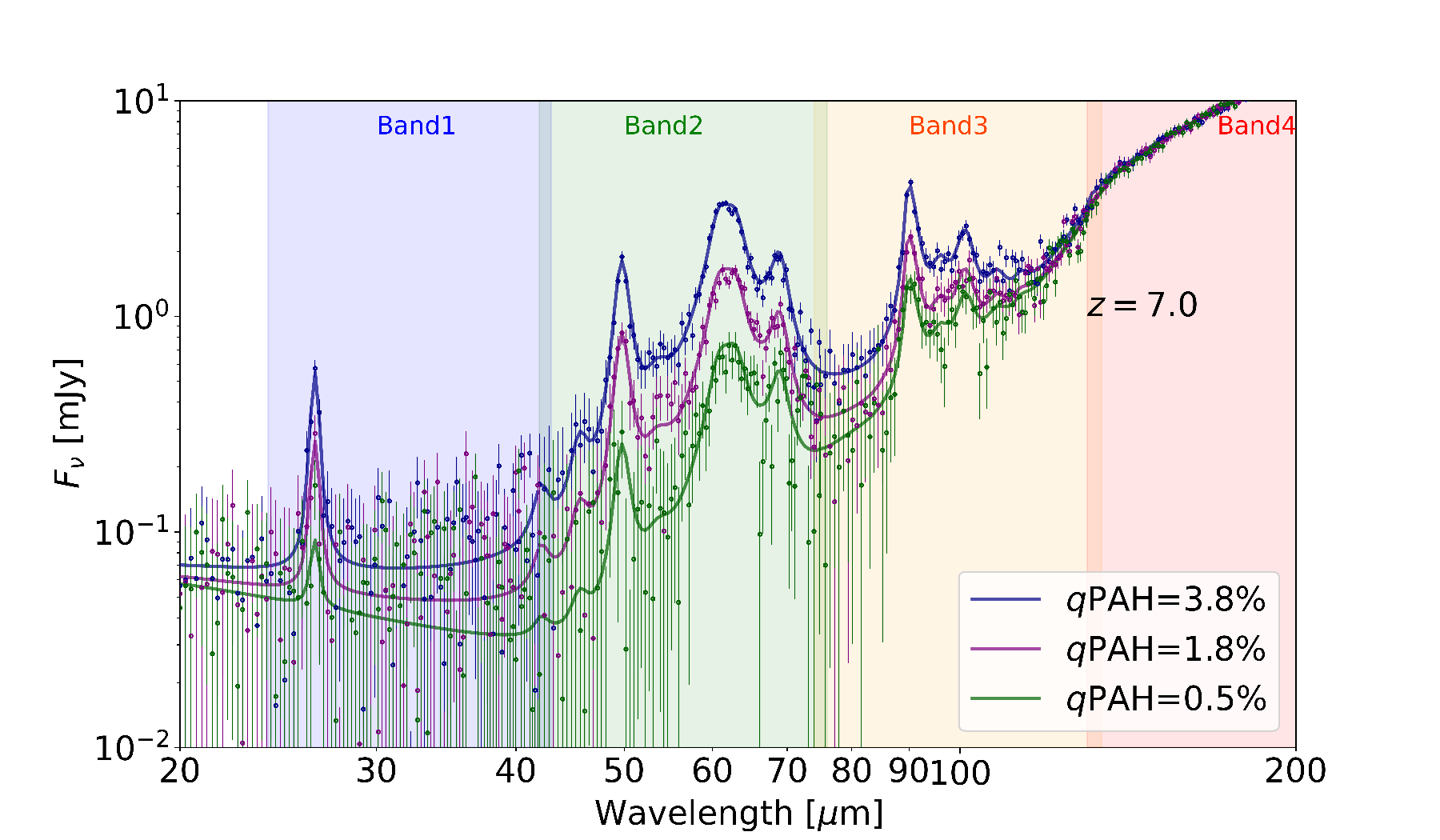} &
\includegraphics[width=0.49\textwidth]{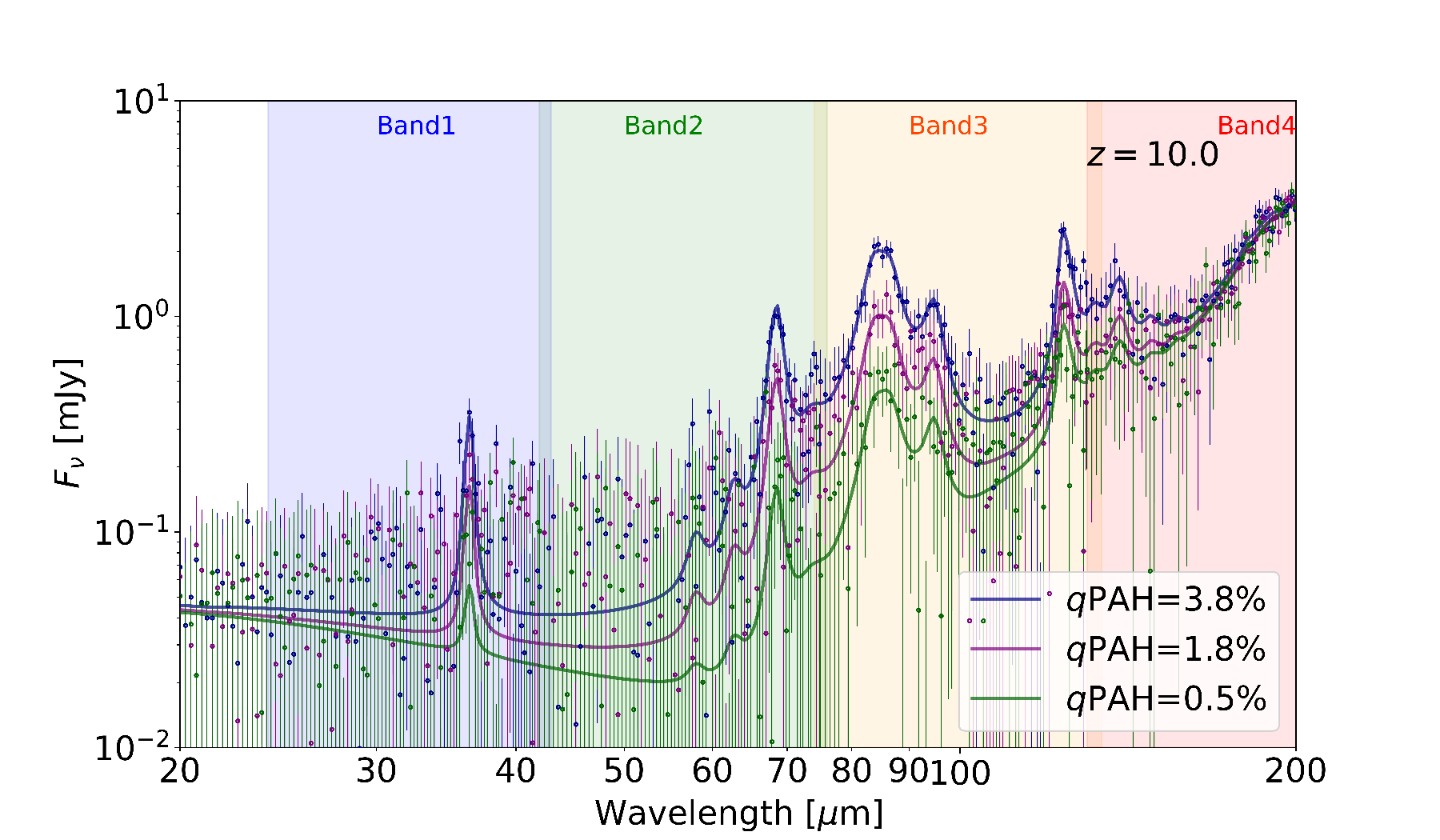}\\
\scriptsize (c) $z=7.0$ & \scriptsize (d) $z=10.0$
\end{tabular}
\end{center}
\caption 
{Simulated PRIMA/FIRESS spectrum of the PAH emission based on the model spectrum for a galaxy with log($L_{\mbox{\tiny IR}}/L_{\odot}$)$=13.6$ in Fig.\ref{fig:firess_redshift}(b) for selective redshifts ($z=3, 5, 7, 10$). In addition to $q$PAH$=0.038$, two lower $q$PAH values (1.8 and 0.5\%) are used for simulation. Dark blue, purple, and green solid lines represent the noiseless model spectrum with $q$PAH$=0.038, 0.018, 0.005$ respectively.}\label{fig:firess_spec_simulation}
\end{figure}

For investigating the science case of the PRIMA/FIRESS to observe PAH emission spectrum from high-redshift galaxies, we simulate the noise-added spectrum with the spectral channel width specified in Table~\ref{tab:instpar}. A $1\sigma$ Gaussian noise based on the $5\sigma$ point source sensitivity curve with 1-hour integration (thick gray dashed lines in Fig.~\ref{fig:pah_model_spec_illustration}(b)) is added to the noiseless model spectrum. Additionally, we convolve the model spectrum with the PRIMA/FIRESS spectral response function (i.e., Gaussian with a finite FWHM\footnote{FWHM for the spectral response function is specified per spectral channel with very small difference between adjacent channels. However, for simplicity, we use a single averaged value for each band: 0.186, 0.427, 0.761, and 1.32$\mu$m for Band 1, 2, 3, and 4, respectively.} from Bradford et al. (this volume)) to consider a proper instrumental effect. Here we assume that the noise due to the PRIMA/FIRESS point source sensitivity is larger than a systematic noise due to the effect of slit width (less significant for a point source) and the background flux confusion noise primarily due to PRIMA's small (1.8m) mirror diameter is less significant for the FIRESS spectroscopy because of an additional velocity dimension as discussed in the conceptual study of the previous FIR spectrometer mission, Origins Space Telescope\cite{bonato_etal_2019}. All synthetic spectra (Fig.~\ref{fig:firess_spec_simulation}) presented in Section~\ref{sec:pah_measurement} are the simulations of PAH emission from a model galaxy with log($L_{\mbox{\tiny IR}}/L_{\odot}$)$=13.6$ (Fig.~\ref{fig:firess_redshift}(b)) observed by PRIMA/FIRESS with 1-hour integration.

\section{Result}
\label{sec:result}
This study characterizes the capability of the FIRESS low-resolution spectroscopy for detecting PAH emission from high-redshift galaxies and for measuring the luminosity of PAH bands.
Following the methods described in Section~\ref{sec:method}, we simulate the spectra composed of PAHs, dust, and stellar radiation observed by PRIMA/FIRESS for a 1-hour long integration. Although the lognormal column density distribution with a power law tail produces a significantly brighter spectrum than the one from the lognormal distribution, for a conservative result, we choose a lognormal $N_{\mbox{\tiny H}}$ distribution. Before we present the result, we note two important assumptions in this investigation.

First, we use a point source sensitivity of the PRIMA/FIRES in this investigation, which implies that the sources are not spatially resolved. The pixel size of the PRIMA/FIRESS specified in Table~\ref{tab:instpar} ranges from 7.6\arcsec\ in Band 1 to 22.9\arcsec\ in Band 4. Therefore, the adopted point source sensitivity works ideally for isolated high-redshift galaxies without close neighbors within 70-140 kpc distance at $z=2\sim4$. However, for spatially blended sources, the redshift information from spectroscopic observation may help alleviate the spatial confusion as discussed in Section~\ref{sec:simul_firess_spec}. Second, we assume an emitting region with 2 kpc diameter ($\ell=2$kpc) when creating an ensemble-averaged spectrum. We chose 2 kpc diameter (or 1 kpc radius) because the redshift evolution of galaxy size based on a large ($\sim$190,000) sample of galaxies after morphological $K$-correction suggests that the effective radii of galaxies at $z=0-6$ are approximately 1 kpc\cite{shibuya_etal_2015}. However, recent observations suggest that the effective radii of galaxies at very high-redshift ($z\sim10$) discovered by JWST are typically a few 100 pc\cite{morishita_etal_2024}, although the stellar mass and size relation from a small number of samples at such high-redshift is not well characterized yet.

If an effective radius of 300 pc is assumed for dust in galaxies at $z\gtrsim 7$, one can infer from Equation~\ref{eq:pahspec} that the overall brightness of the simulated spectra decreases by a factor of 10 and the required integration time for detection with the same $5\sigma$ significance becomes 100 hours. However, we note that even under this assumption (i.e., 300 pc effective radius), the peak flux density of the PAH emission bands at 6.2, 7.7, and 11.2$\mu$m for IR bright galaxy at $z=7$ (log($L_{\mbox{\tiny IR}}/L_{\odot}$)$=13.6$ shown in Fig.~\ref{fig:firess_redshift}(b)) are larger than $5\sigma$ sensitivity with 10-hour integration and small galaxies at the end of reionization can be detected by PRIMA/FIRESS with modest observing time ($\gtrsim 10$ hours).

\subsection{PRIMA FIRESS observing bands and sensitivities for PAH detection}
\label{sec:detection_redshift}
We simulate the noiseless PAH spectra observable with the FIRESS observing bands in different redshifts and compare them with the $5\sigma$ sensitivities. Fig.~\ref{fig:firess_redshift} shows the simulated PAH spectra with $q$PAH$=0.038$ at $z=2,3,5,7,10$ for two different lognormal $N_{\mbox{\tiny H}}$ distributions that have the same variance but different means (Panel (a) and (b)). The purple dashed line and purple dot-dashed line indicate $5\sigma$ point source sensitivity from 1-hour and 10-hour integration, respectively. As described in Section~\ref{sec:pah_spec_dist}, a chosen lognormal distribution determines the distribution of $U$ and the total IR luminosity (annotated in each panel). 

First, one can see that 7.7$\mu$m PAH emission band moves into the observing window of Band 1 for redshift $z=2\sim3$ (cosmic noon), Band 2 for redshift $z=5\sim7$ (epoch of reionization), and Band 3 for redshift $z>7$ (cosmic dawn). Second, most PAH bands are observable up to $z\sim7$ for a spectrum based on a lognormal distribution with $\bar{N}_{\mbox{\tiny H}}\geq1\times10^{23}$cm$^{-2}$ (log($L_{\mbox{\tiny IR}}/L_{\odot}$)$\gtrsim13.6$) although the significance of the observation is weak if $q$PAH is smaller. Third, as shown in Panel(a) for log($L_{\mbox{\tiny IR}}/L_{\odot}$)$=12.13$, the peak of the 6.2, 7.7, and 11.2 $\mu$m band PAH emission is above a $5\sigma$ sensitivity at $z=3$ and 7.7$\mu$m band PAH emission is above a $2.8\sigma$ sensitivity even at $z=5$, after 1-hour integration, which is more than an order of magnitude smaller than the VLA observing time ($10\sim70$ hours as shown in Fig.~\ref{fig:co_pah_lir}(c)) for the CO(1-0) brightness for the similar $L_{\mbox{\tiny IR}}$ (log($L_{\mbox{\tiny IR}}/L_{\odot}$)$\sim12$).

Although the methods of flux measurement and the $q$PAH value may change the significance of detection, the PRIMA/FIRESS with the current instrument specification (Table~\ref{tab:instpar}) is capable of observing PAH emission over a wide range of redshift: from cosmic noon to cosmic dawn if a PAH emitting galaxy is extremely IR bright (HyLIRGs or lensed star-burst galaxies with $L_{\mbox{\tiny IR}}\gtrsim10^{13} L_{\odot}$) and has a dense ($\bar{N}_{\mbox{\tiny H}}\sim1\times10^{23}$cm$^{-2}$) ISM. If the integration time increases to 10 hours (i.e., $3.3\times$ more sensitive than 1-hour integration), the galaxy $L_{\mbox{\tiny IR}}$ limit to detect PAH emissions becomes lower and the ULIRGs-like ($L_{\mbox{\tiny IR}}\gtrsim10^{12} L_{\odot}$) objects can be detectable in 7.7$\mu$m PAH band up to the end of cosmic reionization epoch ($z\sim5$) as seen in Fig.~\ref{fig:firess_redshift}(a).   

\begin{figure}
\begin{center}
\begin{tabular}{cc}
\includegraphics[width=0.49\textwidth]{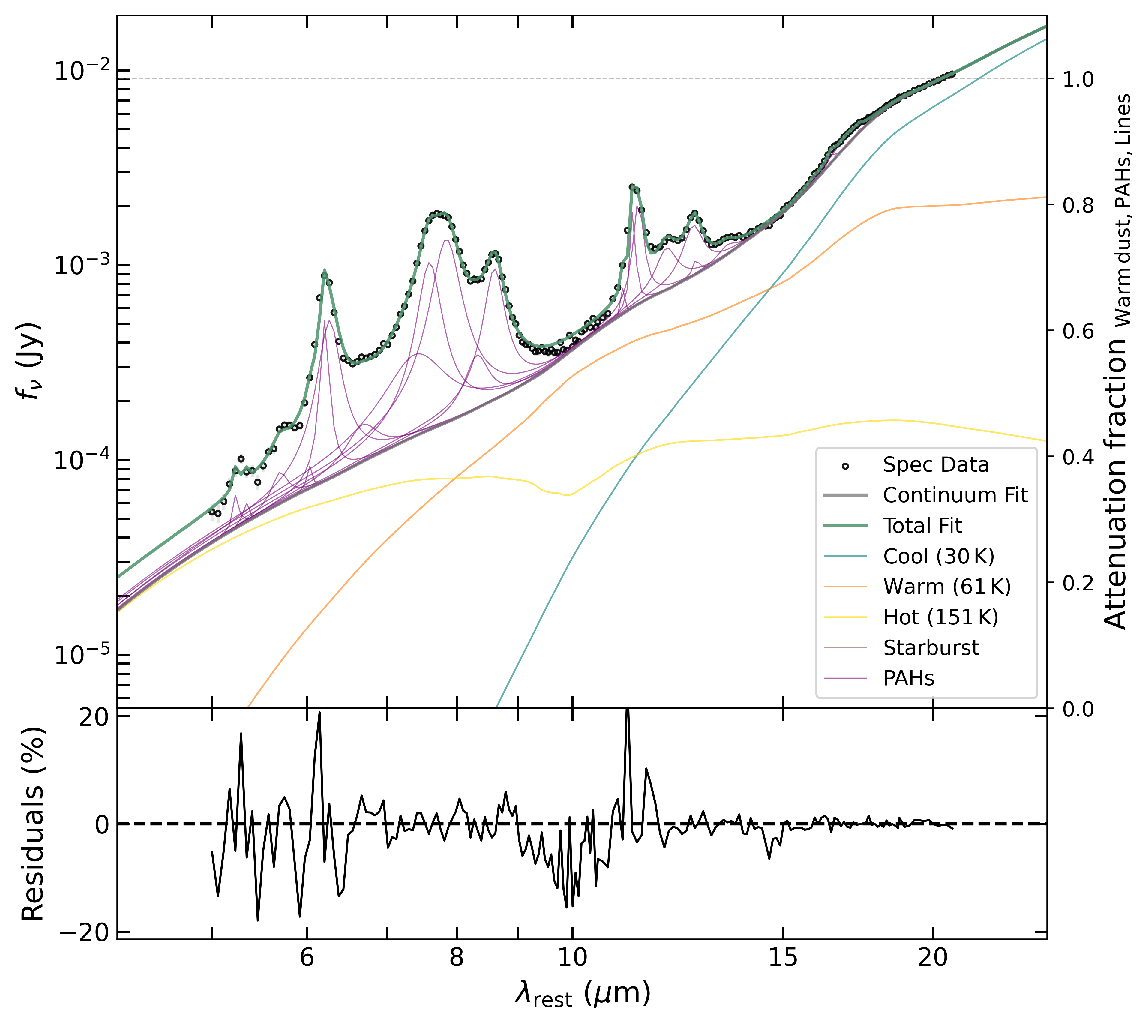} &
\includegraphics[width=0.49\textwidth]{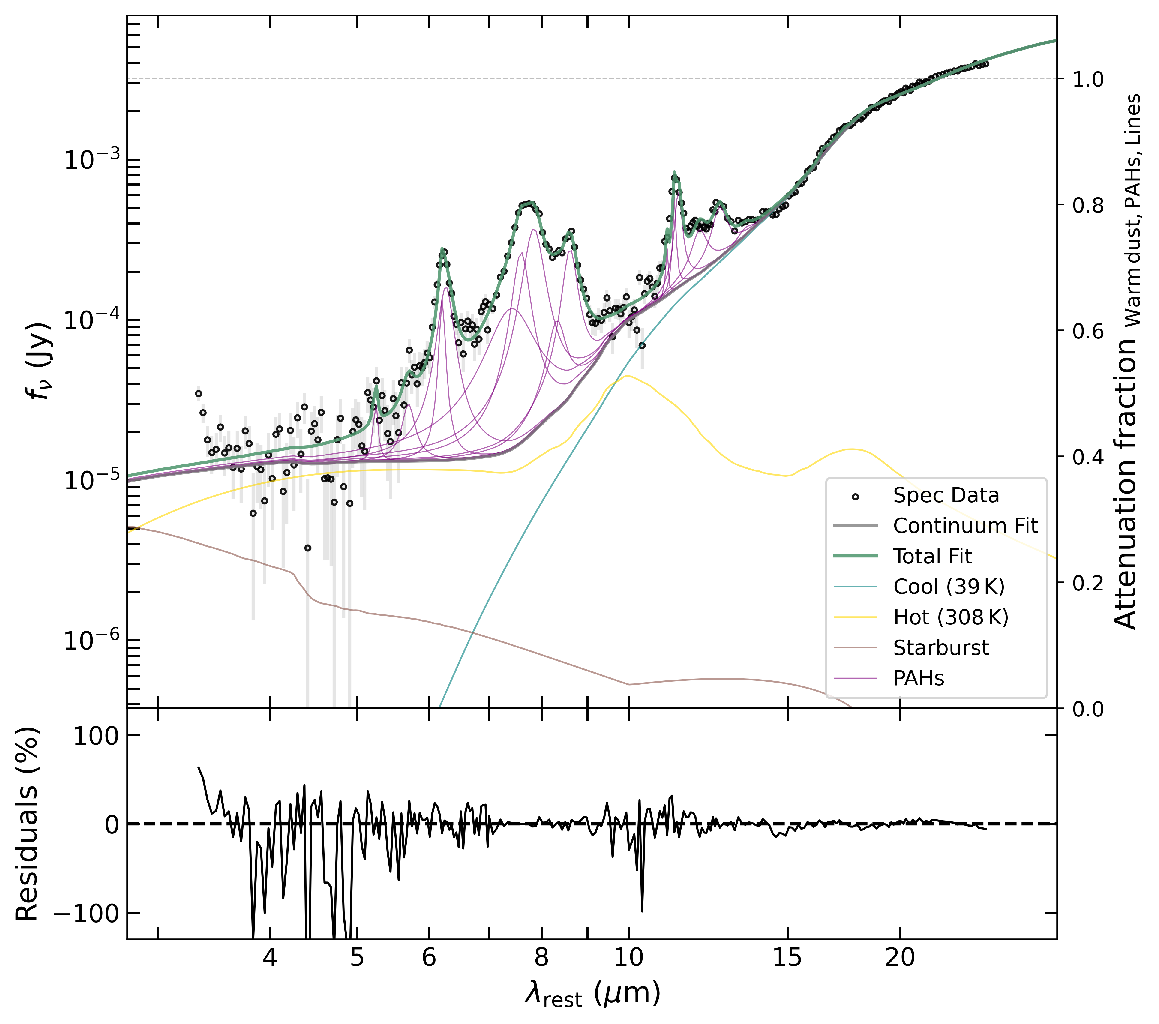}\\
\scriptsize (a) $z=3.0$ with $q$PAH$=0.018$ & \scriptsize (b) $z=5.0$ with $q$PAH$=0.018$\\
\includegraphics[width=0.49\textwidth]{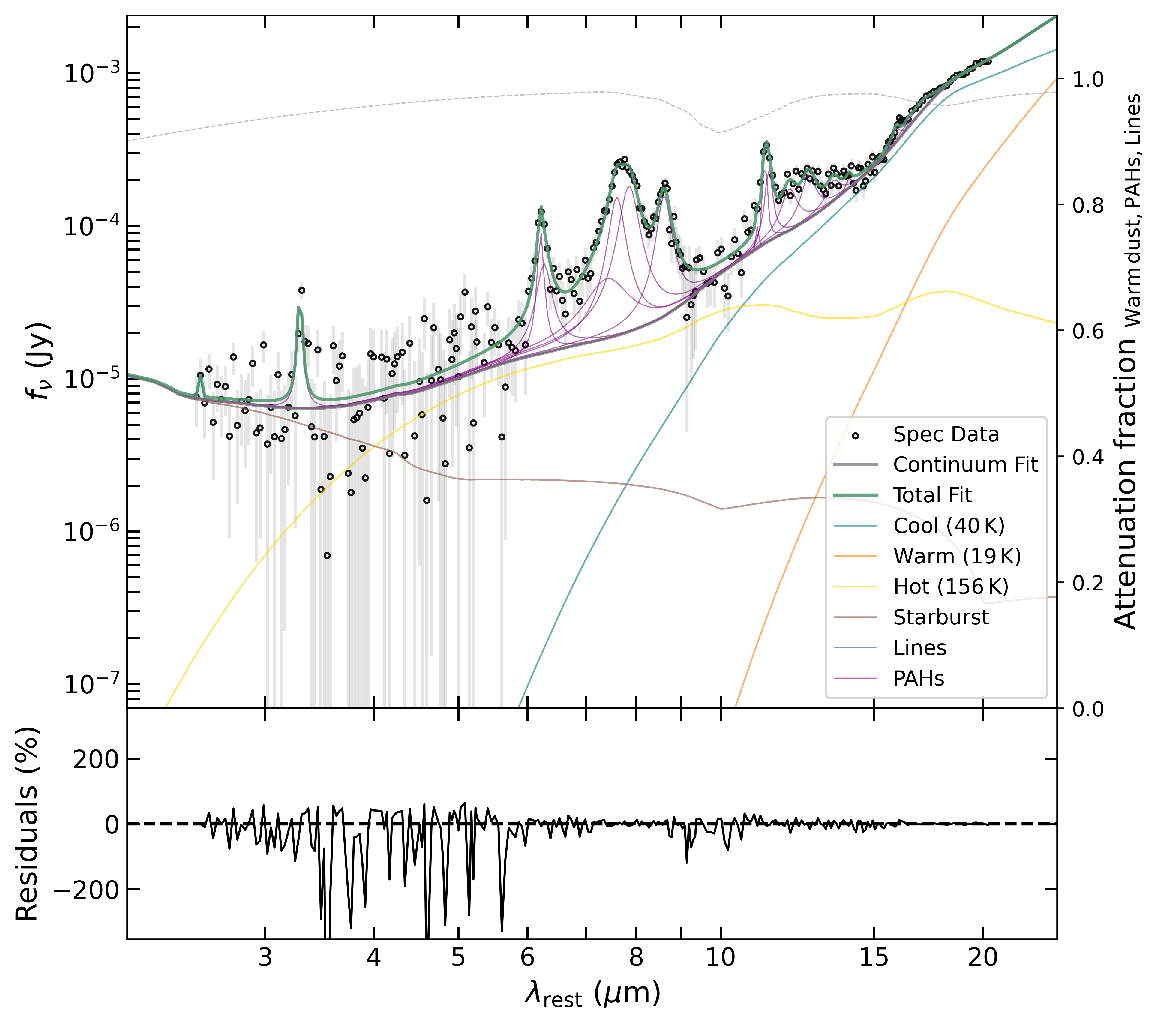} &
\includegraphics[width=0.49\textwidth]{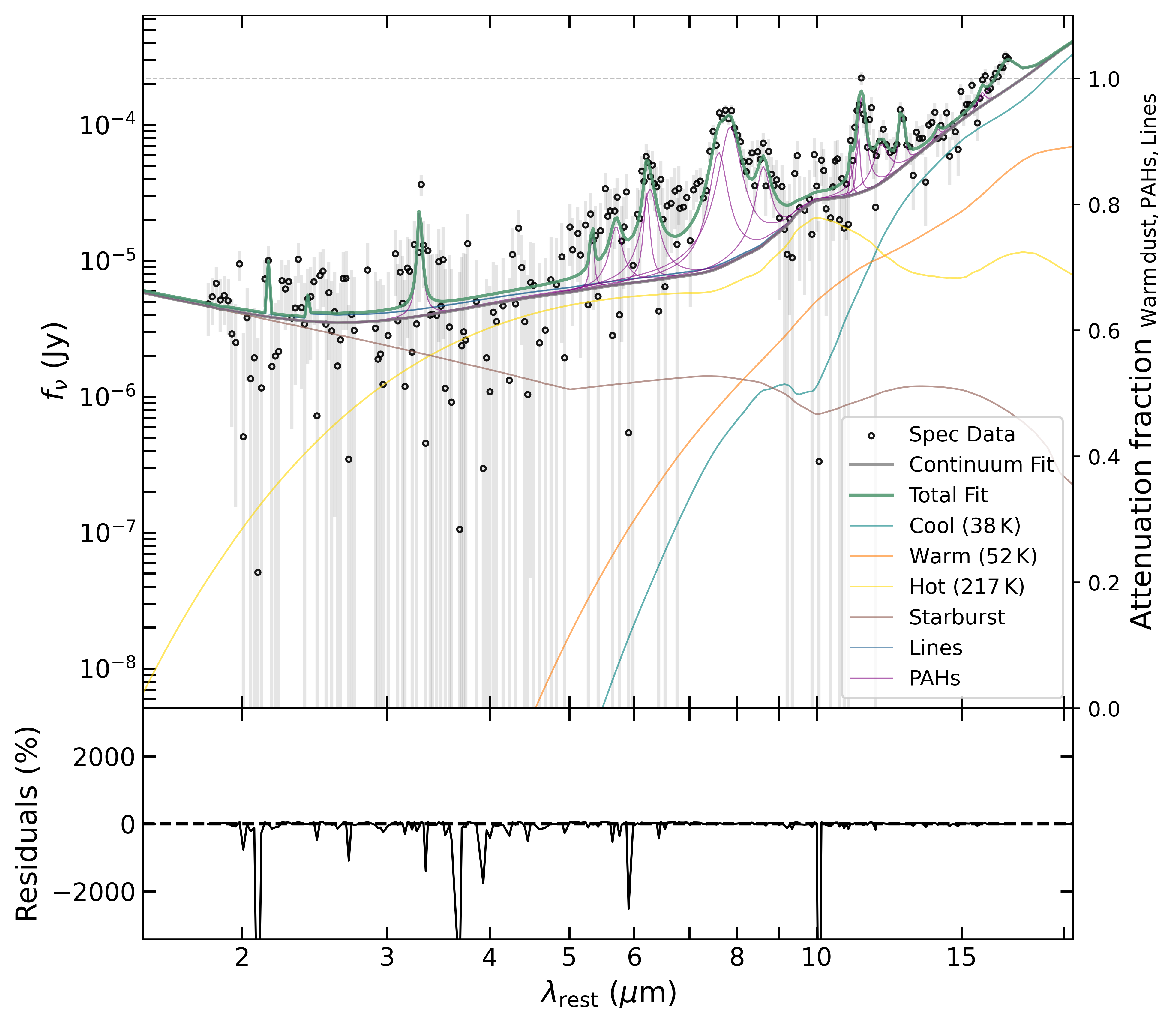}\\
\scriptsize (c) $z=7.0$ with $q$PAH$=0.018$ & \scriptsize (d) $z=10.0$ with $q$PAH$=0.018$
\end{tabular}
\end{center}
\caption 
{Full spectral modeling of the PAH spectrum using CAFE. The spectral data (black circle) with gray error bars in each panel for different redshifts is the simulated PAH spectrum with $q$PAH$=0.018$ in the corresponding panel in Fig.~\ref{fig:firess_spec_simulation}. The attenuation of the model PAH emission (gray dashed line showing attenuation fraction of 1.0) is negligible except for Panel (c) showing a mild variation of the attenuation (10\% at maximum).}\label{fig:pah_measure}
\end{figure}

\subsection{PAH flux measurement from PRIMA/FIRESS spectra}
\label{sec:pah_measurement}
In Fig.~\ref{fig:firess_spec_simulation}, we show the noise-added spectra (explained in Section~\ref{sec:simul_firess_spec}) for a galaxy observed at $z=3,5,7,10$ simulated from the noiseless model spectra in Fig.~\ref{fig:firess_redshift}(b) with the spectral channel width and sensitivity specified in Table~\ref{tab:instpar}. In each panel, the three spectra represent the simulation with high, intermediate, and low $q$PAH values (explained in Section~\ref{sec:pah_spec_qpah}) shown by dark blue (3.8\%), purple (1.8\%), and green (0.5\%) colors. The spectrum becomes noisy due to decreasing flux density with increasing redshift. To measure the flux density in the PAH emission bands in Fig.~\ref{fig:firess_spec_simulation}, we use the PAH spectral modeling tool, CAFE\footnote{\linkable{https://github.com/GOALS-survey/CAFE}} that was originally developed for fitting Spitzer/IRS spectra and later updated to work with JWST IFU data.

Fig.~\ref{fig:pah_measure} shows the result of the CAFE model fitting to the spectral data. Each panel shows the spectrum for intermediate $q$PAH value ($q$PAH$=0.018$) and the best-fit CAFE model, with the residual spectrum in the bottom panel. The input parameter file for starburst galaxies is used for modeling the PAH spectrum for all four cases. In each panel, multiple PAH complexes are shown in purple lines with additional baseline emissions including emissions from cool, warm, and hot dust components, some of which may be degenerate (e.g., warm and cool dust temperatures are swapped as seen in Fig.~\ref{fig:pah_measure}(c)). The attenuation of the model PAH emissions (gray dashed line in each panel showing the attenuation fraction) for fitting the spectra is negligible (attenuation fraction: 1.0) except for the mild variation of attenuation ($<10$\%) seen in Fig.~\ref{fig:pah_measure}(c). The fits are successful with reasonable residuals: the difference between the data and the best-fit model for the largest deviant data point is $\pm20$\% for $z=3$ and $\pm100$\% for $z=5$ (due to the data points with low signal-to-noise ratio) without a significant systematic bias. Even for lower signal-to-noise ratio spectra at higher redshift ($z=7, 10$), the CAFE produces a robust fit to the 6.2 and 7.7$\mu$m features. In particular, even though the 3.3$\mu$m features are noisy at high redshift ($z\gtrsim7$), the spectral fitting robustly captures the 6.2, 7.7, and 11.2$\mu$m features (Fig.~\ref{fig:pah_measure}(c) and (d)), which shows a promise for the use of the 6.2 and 7.7$\mu$m feature to detect high-redshift galaxies.

We also note that, since the PAH emission spectral features are broad, the high-redshift PAH emission spectrum observed in a single FIRESS band may not have sufficient continuum baseline to determine the PAH flux density accurately. For example, Fig.~\ref{fig:firess_spec_simulation}(a) shows that the FIRESS band 1 does not fully cover 6.2 and 7.7$\mu$m PAH emission from a $z=3.0$ galaxy. Likewise, Fig.~\ref{fig:firess_spec_simulation}(b) shows that the FIRESS band 2 does not fully cover 7.7 and 11.2$\mu$m PAH emission from a $z=5.0$ galaxy. The limited continuum baseline from a single band observation may introduce a bias in the PAH luminosity measurement from spectral fitting or may not even allow a simple measurement based on a flux `clip' using a pre-fixed frequency range\cite{draine_etal_2021}. For example, we find that the model PAH luminosities at 7.7 and 11.2$\mu$m from CAFE using the spectra from all four bands and from only Band 2 for a $z=5.0$ galaxy (Fig.~\ref{fig:firess_spec_simulation}(b)) differ by 4\% for the 7.7$\mu$m PAH and 15\% for the 11.2$\mu$m PAH. Given that the redshifted PAH emission spectrum may cross the edges of the FIRESS bands, the best practice for observing high-redshift PAH emission using PRIMA/FIRESS is probably to use at least two bands, or all bands if possible. 

\begin{figure}
\begin{center}
\begin{tabular}{cc}
\includegraphics[width=0.49\textwidth]{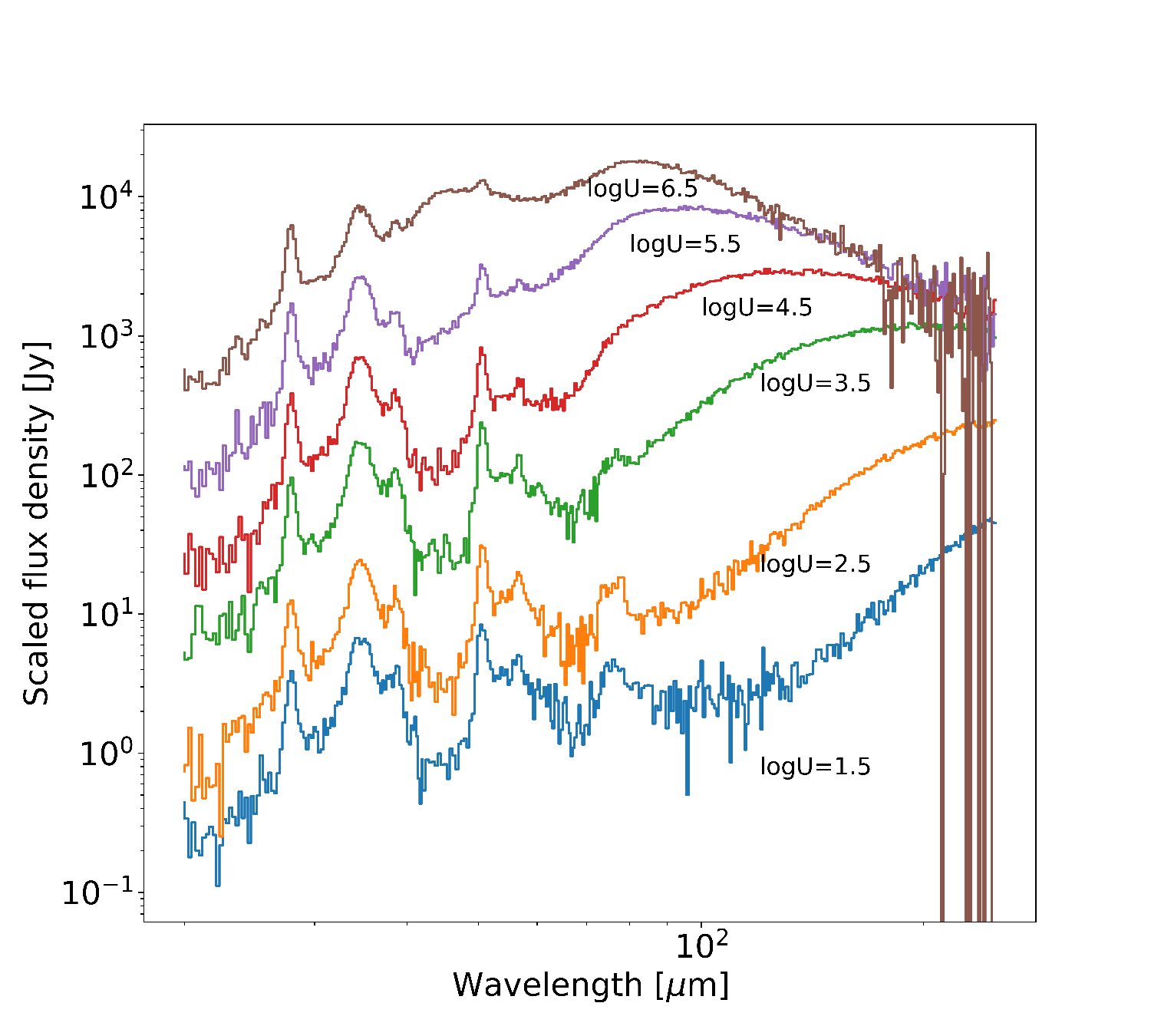} &
\includegraphics[width=0.49\textwidth]{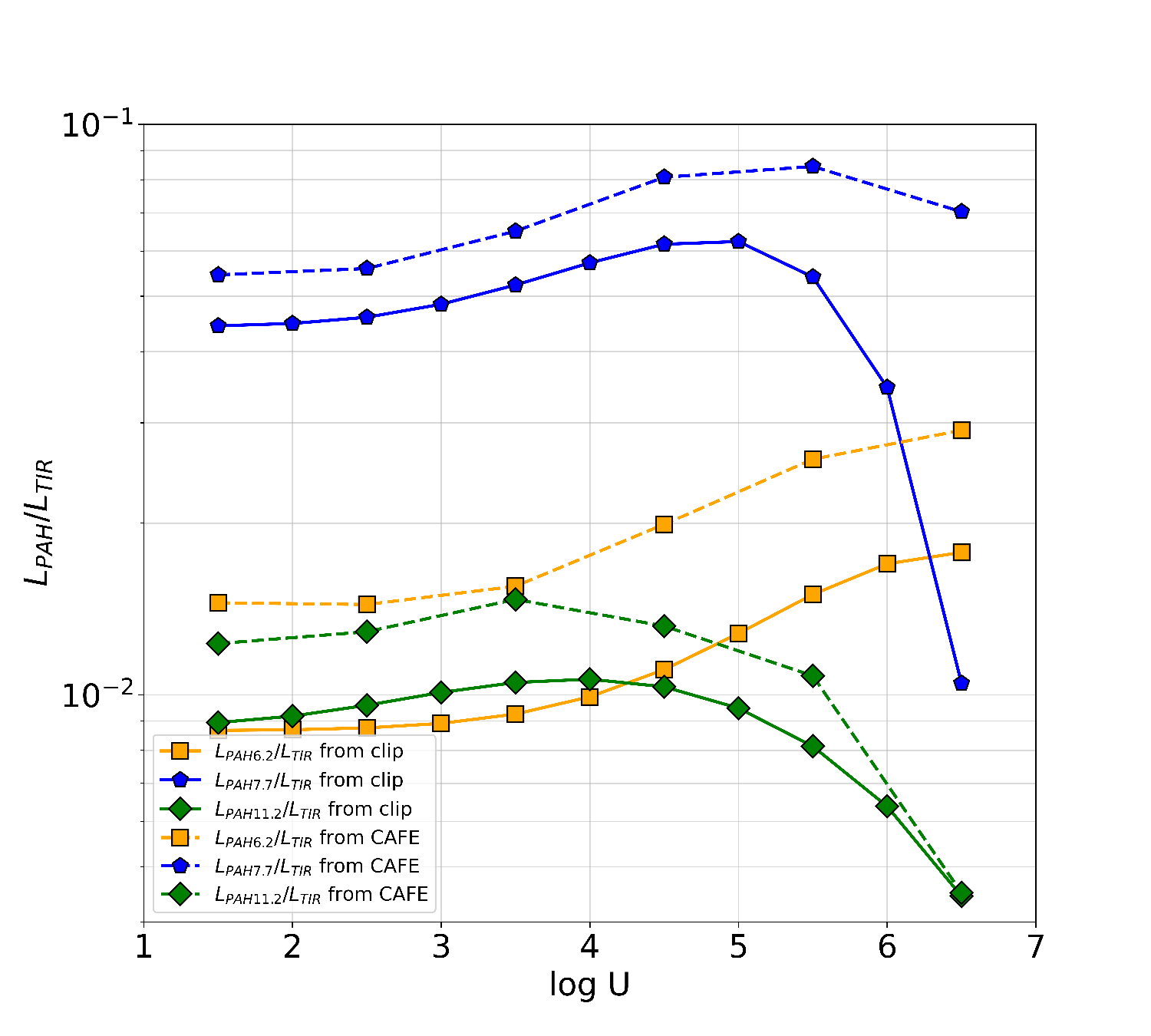}\\
\scriptsize (a) Simulated spectra at $z=3.5$ & \scriptsize (b) PAH luminosity fraction
\end{tabular}
\end{center}
\caption 
{Two different measures of the PAH luminosity (full spectral modeling and `clip' method\cite{draine_etal_2021}) at 6.2, 7.7, and 11.2$\mu$m are compared. {\textit{Panel(a):}} simulated noised added spectra at $z=3.5$ from the model PAH emissivity for given $U$ are shown. The spectra are scaled such that they all have a similar signal-to-noise ratio. {\textit{Panel(b):}} fractional luminosity of the PAH emission relative to $L_{\mbox{\tiny IR}}$ are shown as a function of the ISRF scaling parameter, $U$. The symbols connected by solid lines are obtained by measuring the flux density of the `noise-less' model spectrum using the `clip' method while the symbols connected by dashed lines are obtained by measuring the PAH emission from the spectra in Panel(a) using full spectral modeling tool CAFE.}\label{fig:pah_frac}
\end{figure}

\subsection{Full spectral modeling and `clip' methods}
\label{sec:pah_measurement_comp}
If a single band is used for observation without a sufficient continuum baseline for spectral fitting, one can measure the flux density by integrating the spectrum over a certain frequency range that was originally introduced as `clip' methods\cite{draine_etal_2021}. If one assumes that the physically motivated spectral modeling measures the true luminosities of PAH emission, it is useful to check how close this simple `clipped' flux is to the one obtained by spectral fitting. 

Compared to the full spectral modeling, the flux measured by `clip' methods is expected to be small. However, if the difference between the `clipped' flux and the spectral model flux is just a constant offset and does not depend on the physical conditions of the PAH grains and ISM, the `clip' method is a simple and useful way of understanding the physics of PAH emission (grain size, ionization, ISRF) as often characterized by the PAH band ratios\cite{draine_etal_2021}.

Fig.~\ref{fig:pah_frac}(a) shows the simulated noise-added spectra based on the model PAH emissivity (see Fig.~\ref{fig:pah_model_spec_illustration}(a)) for different $U$ values. The spectra being set to have a similar signal-to-noise ratio are shown with an arbitrary amplitude scale to present them well-separated in Fig.~\ref{fig:pah_frac}(a) for clear visibility. Fig.~\ref{fig:pah_frac}(b) shows the fractional luminosity to the total IR luminosity for the PAH emissions at 6.2, 7.7, and 11.2$\mu$m bands as a function of $U$. The symbols connected by the solid lines represent the value estimated from `clip' methods based on the noiseless model spectra\cite{draine_etal_2021} for given $U$. The same symbols connected by the dashed line represent the value determined by the model PAH emission component from the CAFE spectral modeling of the noise-added simulated spectrum for selective values of $U$ (Panel(a)). For a fair comparison to the `clip' methods which cannot estimate the attenuation corrected intrinsic flux value, we used the value uncorrected for attenuation (i.e., the observed value) from the CAFE model, although the attenuation fractions required for fitting were indeed negligible.\footnote{The attenuation model of PAH spectra based on the 9.7$\mu$m silicate dust optical depth $\tau_{9.7}$ suggests a systematic change of the PAH band ratio as a function of $\tau_{9.7}$\cite{lai_etal_2024}. However, given the relation between A$_V$ and $\tau_{9.7}$ (A$_V$$/\tau_{9.7}=18.5$) for diffuse Galactic dust\cite{mathis_1990}, we note that the A$_V$ value inferred from $N_{\mbox{\tiny H}}$ value\cite{draine_2011}  ($N_{\mbox{\tiny H}}/\mbox{A}_V=1.9\times10^{21}$cm$^{-2}$mag$^{-1}$ for $R_V=3.1$) used in this study results in $\tau_{9.7}\approx1$ causing a relatively small impact.} The same input parameter file for starburst is used for all spectra for the CAFE model except for the spectrum with the highest $U$ value (log$U=6.5$) for which we have to use the input parameter file for AGN because the cool component of dust grains does not produce a good fit for both the continuum flux presented shortward of 3.3$\mu$m PAH feature and a characteristic `AGN-dominated' MIR feature\cite{kirkpatrick_etal_2015}.  

With increasing $U$, a significant change in the spectral shape is noticeable at the $\approx 10\mu$m silicate feature between 7.7 and 11.2$\mu$m PAH features (Fig.~\ref{fig:pah_frac}(a)). As a result, the spectral fitting with different shapes of continuum emission alters the contribution from PAH emission which may result in a significantly different value from the simple estimate based on `clip' methods. In Fig.~\ref{fig:pah_frac}(b), we can see that the fractional luminosity of PAH emission for 7.7 (blue color) and 11.2$\mu$m (green color) measured from the spectral fitting and `clip' methods are in parallel with a constant offset for low $U$ (log$U$$<4$) and then become different from each other for high $U$ (log$U$$\gtrsim4$) when strong ISRF significantly alters the spectral shape of the model PAH emissivity. Unlike the 7.7 and 11.2$\mu$m PAH bands close to the 9.7$\mu$m silicate feature, the 6.2$\mu$m PAH emission (yellow color) shows the same trend of the fractional luminosity as a function of $U$ for the estimate from both spectral fitting and `clip' methods.  

\section{Synergy with Fast Mapping Speed}
\label{sec:pah_mapping}
In this paper, the focus of our investigation is the capability of PRIMA/FIRESS to observe the PAH emissions from high-redshift galaxies using a pointing mode observation. However, PRIMA is also efficient in mapping. The time to survey 100 arcmin$^2$ area for $5\sigma$ line sensitivity, $3\times10^{-19}$W m$^{-2}$, is 800 hours for 23-75$\mu$m (Band 1 and 2) and 336 hours$\times(\lambda/100\mu\mbox{m})^{-1.68}$ for 75-235$\mu$m (Band 3 and 4)\cite{bradford_etal_2024}. This is more than $10\times$ faster than JWST/MIRI and Spitzer/IRS at the similar observing wavelength\cite{bradford_etal_2024}. The nominal sensitivity for pointing observation used in this study ($1.9\times10^{-19}$W m$^{-2}$) is comparable to this mapping sensitivity and therefore our findings on the capability of PRIMA/FIRESS for detecting PAH emission from high-redshift galaxies also hold for PAH mapping (with 63\% degraded sensitivity for the same observing parameters and physical properties of galaxies). One should note that this synergy between the high sensitivity for PAH detection and the fast mapping speed makes PRIMA/FIRESS an efficient instrument for PAH mapping. Here we discuss galaxy protocluster as an example science case for PAH mapping.

High-redshift protoclusters are important tracers of the structure formation in the early Universe. However, it is difficult to spectroscopically confirm the membership of each galaxy in the protocluster due to the large sky area and the faint line emission. If the PAH emission is bright, one can use PRIMA/FIRESS to map the high-redshift galaxy protoclusters with a typical angular size of a few arcmin at $z=2$\cite{muldrew_etal_2015}, which will detect PAH emissions from individual galaxies with sufficient SFR (SFR$=100\sim1000$ M$_{\odot}$yr$^{-1}$ based on the SFR-$L_{\mbox{\tiny IR}}$ correlation\cite{calzetti_2013}, SFR$=10^{-10}(L_{\mbox{\tiny IR}}/L_{\odot})$ M$_{\odot}$yr$^{-1}$) and confirm their redshift. If we use the measured overdensities of the galaxies in protoclusters at $z=2\sim3$ and count their spectroscopically confirmed member galaxies\cite{liu_etal_2025}, the estimated number density of the protocluster member galaxies is $\approx 4$ per arcmin$^2$ (i.e., $\approx 20$\arcsec\ distance to the closest neighbor for each member galaxy). With 7.6\arcsec\ pixel size for Band 1 of the PRIMA/FIRESS in which the observed PAH emissions from galaxies at $z=2\sim3$ are detectable, the PRIMA/FIRESS is capable of observing individual member galaxies without a significant impact of the blended sources in the pixel.

For a galaxy protocluster at cosmic noon ($z=2\sim3$) traced by enhanced number of sub-millimeter galaxies with SFR$=100\sim1000$ M$_{\odot}$yr$^{-1}$, the PRIMA/FIRESS Band 1 can map the entire protocluster and detect PAH emission from its member galaxies (see Fig.~\ref{fig:firess_redshift}) with $\lesssim100$ hours. The membership confirmation will be more reliable by full spectroscopic mapping, for example, using FIRESS Band 1 and 2 to detect multiple PAH emission lines as shown in Fig.~\ref{fig:firess_spec_simulation}(a). PRIMA/FIRESS mapping is much more efficient than observing CO from a handful of selected member galaxies at $z\sim3$ with $\approx30$ hours of VLA time for each galaxy (Fig.~\ref{fig:co_pah_lir}(c)).

\section{Summary}
\label{sec:summary}
In this paper, we perform a simulation of the observed PAH spectra of spatially unresolved high-redshift galaxies for PRIMA/FIRESS with low-resolution ($R\sim100$) spectroscopy. The model PAH emissivity spectrum is converted to the observed spectrum at a given redshift with the FIRESS channel resolution, using cylindrical geometry with column density distribution and ``low,'' ``intermediate,'' and ``high'' $q$PAH values. Random Gaussian noise added to the spectra is created based on the PRIMA/FIRESS sensitivity and the PRIMA/FIRESS spectral response function is convolved with the model spectra. The principal results of this study are as follows.

\begin{enumerate}
\item We calculate the observed PAH spectra based on an ensemble distribution of gas column density following the form of lognormal and lognormal with a power-law tail. The addition of a small fraction of power-law tail that represents high-density star-forming regions with strong ISRF, changes the shape of the averaged spectrum and increases its brightness.   
\item The bright 7.7$\mu$m PAH emission band is shifted to the PRIMA/FIRESS Band 1 window for $z=2\sim3$, Band 2 window for $z=5\sim7$, and Band 3 window for $z>7$. For 1 hour integration, most of the well-known PAH bands (3.3, 6.2, 7.7, 11.2$\mu$m) are observable at $z\sim7$ for a galaxy with $L_{\mbox{\tiny IR}}\sim1\times10^{13}L_{\odot}$. For 10 hour integration, the 7.7$\mu$m PAH band is observable at $z\sim5$ for a galaxy with $L_{\mbox{\tiny IR}}\sim1\times10^{12}L_{\odot}$. The required observing time for detecting PAH emission is an order of magnitude lower than the VLA observing time for CO(1-0) observation of a galaxy with a similar $L_{\mbox{\tiny IR}}$.
\item We simulate the measurement of the PAH emission by modeling the synthetic spectra using physically motivated near- and mid-IR spectral modeling tool, CAFE. For a range of redshifts ($z=3,5,7,10$), the CAFE fits the synthetic spectra and produces robust fitting results: for the data point with the  largest deviation from the best-fit model, $\pm20$\% residual for $z=3$ and $\pm100$\% residual for $z=5$ (due to the data points with the lowest signal-to-noise ratio) without a significant systematic bias. Even for lower signal-to-noise ratio spectra at higher redshift ($z=7,10$), the CAFE produces a robust fitting to the 6.2 and 7.7$\mu$m features. For reliable spectral modeling of the broad PAH emission feature, which may cross the edge of the FIRESS band, the best practice is to use at least two bands for observation.
\item We compare the measurement of the PAH luminosity from a simple `clip' method based on a pre-fixed frequency range and from spectral fitting methods. For weak ISRF strength (log$U$$<4$), the fractional luminosities of PAH emissions relative to the total IR luminosity as a function of $U$ shows the same trend for 6.2, 7.7, 11.2$\mu$m, however with increasing ISRF intensity (log$U$$\gtrsim4$), the measurements from the two methods do not show the same trend: the difference is significant in the 7.7 and 11.2$\mu$m features while the 6.2$\mu$m feature still shows the same trend for the measurement from both methods.
\item PRIMA/FIRESS can be used as a mapping instrument to measure star formation and redshift of the member galaxies in the entire region of a galaxy protocluster at $z=2-3$ by observing bright PAH emissions.
\end{enumerate}

Our study suggests that PRIMA/FIRESS low-resolution spectroscopy has promise for observing PAH emissions from high-redshift galaxies. The specification of the FIRESS instrument complements the long-wavelength channels of the JWST/MIRI MRS with 1-2 orders of magnitude better sensitivity. Thus, PRIMA/FIRESS can extend PAH observation to higher redshifts ($z>4$) where the spectroscopic observation of PAH using JWST/MIRI MRS is practically infeasible.

\section* {Code and Data Availability}
The code used to generate the results and figures is available in the following Github repository. Data sharing is not applicable to this article, as no new data were created or analyzed.\\
{\url{https://github.com/ilsangyoon/PRIMA_FIRESS.git}}

\section* {Disclosure}
The authors declare that there are no financial interests, commercial affiliations, or other potential conflicts of interest that could have influenced the objectivity of this research or the writing of this paper.

\section* {Acknowledgments}
The authors thank two anonymous referees for their constructive comments which improved the paper. The National Radio Astronomy Observatory is a facility of the U.S. National Science Foundation operated under cooperative agreement by Associated Universities, Inc. This research was carried out in part at the Jet Propulsion Laboratory, California Institute of Technology, under a contract with the National Aeronautics and Space Administration. I.S. acknowledges fundings from the European Research Council (ERC) DistantDust (Grant No.101117541) and the Atraccíon de Talento Grant No.2022-T1/TIC-20472 of the Comunidad de Madrid, Spain. IGB is supported by the Programa Atracci\'on de Talento Investigador ``C\'esar Nombela'' via grant 2023-T1/TEC-29030 funded by the Community of Madrid.


\bibliography{report}   
\bibliographystyle{spiejour}   


\vspace{2ex}\noindent\textbf{Dr. Ilsang Yoon} is a scientist at the National Radio Astronomy Observatory. He holds a Ph.D. in Astronomy from the University of Massachusetts Amherst. Dr. Yoon's research interests include observational study of galaxies and their supermassive black holes in local and distant Universe using primarily radio, millimeter, and far-infrared observing facilities. He has  expertise in calibrating radio interferometric data and developing data quality assurance heuristics. 



\end{spacing}
\end{document}